\documentclass[twocolumn, superscriptaddress,amsmath,amssymb,twocols,nofootinbib,aps]{revtex4}
\usepackage{verbatim}
\usepackage{graphicx}
\usepackage{dcolumn}
\usepackage{bm}
\usepackage{mathtools}  
\usepackage{booktabs}                              
\usepackage{float}
\usepackage{hyperref}        
\usepackage{multirow}
\usepackage{hhline}                         
\usepackage[usenames, dvipsnames]{color}
\usepackage{color}
\usepackage{xr}
\usepackage{cases}
\usepackage{soul}
\usepackage{lipsum}
\usepackage[makeroom]{cancel}

\newcolumntype{P}[1]{>{\centering\arraybackslash}p{#1}}

\begin{document}

\preprint{APS/123-QED}

\title{Artificial life of an active droplets system: a quantitative lifecycle analysis}

\author{Matteo Scandola}
\affiliation{Laboratoire Jean Perrin, UMR 8237 Sorbonne Université/CNRS,
Institut de Biologie Paris Seine, 4 Place Jussieu, F-75005 Paris, France}
\affiliation{Department of Cellular, Computational and Integrative Biology (CIBIO), University of Trento, via Sommarive 9, I-38123 Trento, Italy}
\author{Silvia Holler}
\affiliation{Department of Cellular, Computational and Integrative Biology (CIBIO), University of Trento, via Sommarive 9, I-38123 Trento, Italy}
\author{Richard J. G. L\"offler}
\affiliation{Centre for Star and Planet Formation, Globe Institute, University of Copenhagen, Copenhagen, Denmark}
\author{Martin M. Hanczyc}
\email{martin.hanczyc@unitn.it}
\affiliation{Department of Cellular, Computational and Integrative Biology (CIBIO), University of Trento, via Sommarive 9, I-38123 Trento, Italy}
\affiliation{Chemical and Biological Engineering, University of New Mexico, Albuquerque, NM, 87106, USA}
\author{Raffaello Potestio}
\email{raffaello.potestio@unitn.it}
\affiliation{Physics Department, University of Trento, via Sommarive, 14 I-38123 Trento, Italy}
\affiliation{INFN-TIFPA, Trento Institute for Fundamental Physics and Applications, I-38123 Trento, Italy}
\author{Roberto Menichetti}
\email{roberto.menichetti@unitn.it}
\affiliation{Physics Department, University of Trento, via Sommarive, 14 I-38123 Trento, Italy}
\affiliation{INFN-TIFPA, Trento Institute for Fundamental Physics and Applications, I-38123 Trento, Italy}

\date{\today}

\begin{abstract}
The study of synthetic active matter systems holds the promise for designing smart materials and devices with emergent characteristics akin to those of living organisms, eventually opening the doors to the realization of artificial life. Such an investigation, however, is challenged by the difficulty inherent in identifying the relationship between the features of the elementary constituents and the emergent properties of the whole; to this end, a key step consists in the accurate quantification of the system's observed behavior. Here, we report the study of $50$ self-propelled oil droplets floating on the surface of an aqueous solution. $25$ droplets are stained with a red dye, and the other $25$ are stained blue: the colorants affect the droplets' interfacial tension properties differently, consequently influencing their collective dynamics. Droplet trajectories extending for up to $5$ hours are extracted from video recordings with a tracking pipeline developed \emph{ad hoc}. The structural and dynamical analysis of the system reveals a ``life-to-death'' cycle unfolding in qualitatively distinct stages, showcasing a complex interplay between individual droplet mobility and collective organization. The tools developed and the results obtained in our work pave the way to the \emph{in silico} modelling as well as the experimental design of synthetic active matter systems displaying life-like and programmable behavior.
\end{abstract}


\maketitle

\section{Introduction}

Active matter encompasses systems composed of self-driven constituents that continuously dissipate energy and use it to move or exert mechanical forces on their environment \cite{gompper20252025}. Such systems are widespread, arising from biological and synthetic origins as well as from a combination of both. In biological contexts, they occur at all possible levels of scales and complexity, ranging from bio-polymers like microtubules and actin \cite{10.1063/5.0011466} (both integral components of the cellular cytoskeleton), to living collectives as cells, bacteria, schools of fish, and flocks of birds \cite{ouellette2022physics}. Synthetic active matter systems, in contrast, are engineered assemblies capable of self-propelled and dynamic behavior reminiscent of those observed in their biological counterparts. Examples include, e.g., light-activated particles, self-propelled droplets, catalytic Janus colloids, and vibrated granular rods \cite{te2025metareview}, all of which convert energy into directed motion. Finally, combining synthetic and biological active components is a recent avenue that could give rise to novel, hybrid behaviors not found in either system alone \cite{chen2025bioinspired}. Irrespective of their origin, a pivotal feature of active matter systems is that, even in the case in which the \emph{individual} constituent entities display a relatively plain morphology, they \emph{emergently} can exhibit extremely complex behavior, characterized by intriguing phenomena such as coherent motion \cite{PhysRevLett.75.1226,kaiser2017flocking}, dynamic self-assembly \cite{wang2015one,hokmabad2022spontaneously}, motility-induced phase separation \cite{bechinger2016active,omar2021phase}, and active turbulence \cite{alert2022active}. Furthermore, from a purely theoretical standpoint, the body of knowledge of equilibrium statistical
physics displays limited applicability in the investigation
of self-propelled systems, rendering active matter an invaluable source---as well as a testbed---for advancing the field of non-equilibrium statistical mechanics \cite{marchetti2013hydrodynamics,lowen2020inertial,fodor2022irreversibility,cavagna2023natural}.

\begin{figure*}[t]
    \centering
    \includegraphics[width = \textwidth]{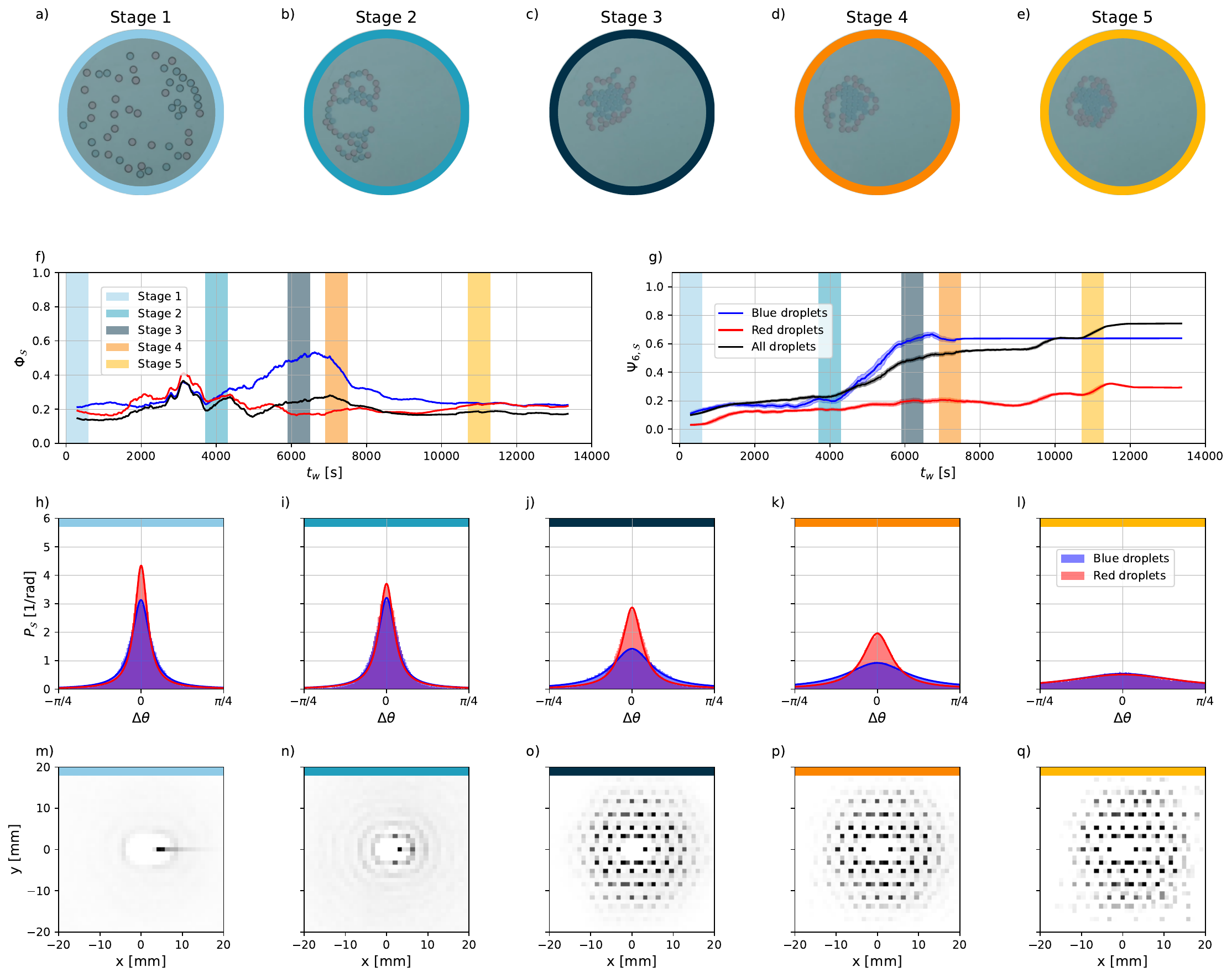}
    \caption{(a-e) Snapshots of the experimental setup, with progression from left to right illustrating typical droplet configurations at the onset of each evolutionary stage of the system. The images are circled with the color of the corresponding stage, as indicated in the key provided in the plots below. (f-g) Time dependence of the droplets' velocity polarization $\Phi_{\mathcal{S}}$ and hexatic order parameter $\Psi_{6,\mathcal{S}}$, respectively defined in Eq.~\ref{eq:order} and~\ref{eq:hex_order} of the main text. Colored vertical bars mark the onset of each evolutionary stage. (h-l) From left to right: turning angle distribution $P_{\mathcal{S}}$, see Eq.~\ref{eq:turn_angl}, of the blue and red droplet populations at the onset of each evolutionary stage. (m-q) From left to right: blue-blue dimer distributions $D_{\mathcal{B}\mathcal{B}}$, see Eq.~\ref{eq:dimer}, at the onset of each evolutionary stage.}
    \label{fig:fig1}
\end{figure*}

Within the broad realm of synthetic active matter, artificial life research focuses on constructing and investigating simplified chemical systems that operate far from equilibrium and showcase life-like properties, with the purpose of shedding light on the physical principles underlying living matter. The system realizations investigated in this field exhibit a wide repertoire of behaviors observed in living organisms, including, besides motion \cite{hanczyc2014droplets, loffler2023ecosystem}, fission and fusion \cite{caschera2013oil}, oscillation \cite{jancik2022swarming, krishna2024dynamic}, chemotaxis \cite{cejkova2014dynamics,holler2018transport, holler2020autoselective}, and phototaxis \cite{suzuki2016phototaxis, zarghami2019dual}. Active droplets are often experimentally employed in this endeavor \cite{hanczyc2014droplets,doi:10.7566/JPSJ.86.101004,loffler2023ecosystem}; compared to other examples within the synthetic active matter domain, however, a characteristic feature of the droplets systems employed in artificial life research is their typical scale \cite{te2025metareview}, as they generally consist of millimeter to centimeter-sized agents, either self-propelling on the surface of an aqueous solution or being immersed in the bulk, moving on the bottom of the container. Their size, along with their relatively simple composition (typically less than five chemical constituents), makes them experimentally very accessible and requires no advanced laboratory equipment. In these systems, the self-propelled motion of the droplets originates from the delicate interplay between the features of their constituents and the surrounding environment: their activity is based on the generation of a chemical gradient and on the solutal Marangoni effect, in which objects release surfactants creating an inhomogeneous concentration field that lowers the surface tension, which in turn generates a bulk convection and surface flow towards areas with higher surface tension. The initial motion is likely caused by random perturbations of the surfactant gradient that are subsequently self-stabilized by the induced motion of the source of the surfactant \cite{LordRayleigh1890,Nakata1997,Nagai2005}. Any mechanism that prevents saturation of the surface by the surfactant, such as evaporation or dissolution, will maintain the gradient and consequently the motion until the system reaches equilibrium. 

In the context of artificial life research, in this study we focus on a droplet system first reported on by Tanaka and coworkers \cite{doi:10.7566/JPSJ.86.101004}, which expresses collective flocking behavior with emergent patterns that change over the course of an experiment and was suggested to be able to model complex biological phenomena. The system is instantiated by multiple droplets, composed of a mix of paraffin oil and the Ethyl Salicylate (ES) active ingredient, floating on an aqueous phase containing another surfactant (typically Sodium Dodecyl Sulfate). While the exact physicochemical mechanism lying at the origin of the dynamics of this system is still not fully understood, it is supposed that the ES released from the droplets likely reacts with the surfactant already present on the surface to form micellar aggregates that disperse in the bulk aqueous phase. This removes a portion of both types of surfactants from the water surface, locally raising surface tension (a portion of the ES is also evaporated from the surface). As a result, complex spatiotemporal gradients develop, where the surface tension is lower than the equilibrium case in the close vicinity of the droplets and higher further away \cite{tanaka2021surfactant}, giving rise to a close-range repulsion and long-range attraction mechanism between the agents. At the same time, the droplets, being slightly denser than the bulk water phase, are deforming the meniscus of the latter, resulting in capillary attraction that competes with the short-range repulsion caused by the Marangoni effect. Finally, being the system closed, the bulk aqueous phase as well as the surrounding atmosphere gradually become saturated with ES; this diminishes the droplet-induced gradients and results in the evolution of cluster formation over time, ending in a ``death'' state in which the droplets are largely steady and arranged in a crystal-like hexagonal pattern.

Experimental systems such as the one described above represent an exciting model platform illustrating the principle of life-like emergence, where unexpected higher order structures come to be from fairly simple interactions in a collective---namely, \emph{``the whole is greater than the sum of its parts''}. While these experiments are \emph{per se} fascinating and visually striking from the intuitive standpoint of emergence, the practical challenge remains to properly quantify their outcomes and extract meaningful statistical information from them, to aid in understanding the underlying physics that lies at the origin of the observed properties. 

The present work aims, through a combined experimental and theoretical effort, to address this challenge while also iterating on the complexity of the experiment proposed by Tanaka \emph{et al.} \cite{doi:10.7566/JPSJ.86.101004}. From a methodological perspective, a robust tracking pipeline is developed that processes raw experimental videos of the self-propelled system to extract accurate trajectories of the droplets’ centers, effectively mapping these hydrodynamic entities onto point-like particles. The subsequent analysis of these trajectories, based on a series of metrics rooted in statistical physics, enables a systematic quantification of the droplets’ dynamical and structural properties at each stage of the system’s evolution, shedding light on their emergent behavior in all the phases that range from artificial life to quiescence---or ``death''. Although here applied to a specific case study, the proposed tracking pipeline and resulting semi-automated workflow (from experimental videos to a quantitative analysis of the agents’ trajectories) is general and adaptable to other synthetic active matter realizations. 

From an applied perspective, several studies indicate that the dyes that are often added to the active droplets to increase their contrast with the background can affect the oil-water interfacial tension, hence influencing the droplets' features---such as the magnitude of their self-propulsion---and behavior \cite{bettered,loeffler2021new,Watanabe2022}. Here, we thus enhance the sophistication of the Ethyl Salicylate active droplet system of Ref.~\cite{doi:10.7566/JPSJ.86.101004} by mixing two different species of agents: ``blue'' droplets, stained with Sudan black B dye; and ``red'' droplets, stained with Oil red O dye. Apart from this difference, the droplets are identical in composition, and the experimental setup remains the same as the one introduced by Tanaka and coworkers \cite{doi:10.7566/JPSJ.86.101004}. More specifically, we consider a system of $50$ self-propelled ES/paraffine droplets, $25$ blue and $25$ red, moving at the top of an SDS solution within a Petri dish. Other experiments have shown that the coexistence of different types of active droplets can lead to new and more complex behaviors compared to the single-component case \cite{Meredith2020,loeffler2021new,Watanabe2022,Mallick2025}, so that even the minimal change in composition considered here could yield interesting phenomena.

Through the detailed analysis of the droplets' trajectories enabled by our tracking pipeline, it is observed that the features of the agents depend on their color and, as a result, the emergent traits of the mixed system change compared to the homogeneous case \cite{doi:10.7566/JPSJ.86.101004}. Specifically, over the course of an experiment (up to six hours), the collective behavior of the droplets is found to pass through a sequence of phases, or stages, each characterized by qualitatively and quantitatively different system-wide dynamic and structural properties. Five snapshots of the experimental setup are presented in Fig.~\ref{fig:fig1}, corresponding to the five main stages we identify in the system's evolution. These range from a disordered, high-activity phase at the onset of the experiment, displaying no organized patters; to an intermediate phase in which medium-sized, transient structures emerge; to, finally, the realization of a persistent arrangement in a quasi-regular assembly of the agents at late times and for vanished activity. Interestingly, due to the distinct features of the two coexisting species of droplets, some of the individual stages are reminiscent of phenomena observed in active Brownian particle systems---e.g., motility-induced phase separation \cite{PhysRevLett.108.235702}. The fact that these stages occur in sequence, further doing so in a reproducible manner across different experiments, makes the whole setup particularly fascinating, and underlines the necessity for a comprehensive and accurate pipeline, as the one proposed here, to track and quantify the collective motion of this synthetic, life-like active matter system. On the one hand, as a first step in the development of more complex realizations of artificial life displaying, e.g., tailored or even programmable properties---presumably first \emph{in silico}, through the construction of dedicated theoretical models, and then practically due to the overwhelmingly large size of the experimental parameter space of this class of systems. On the other hand, to advance our understanding of the physical principles underlying biological life itself.

\section{Materials and methods}
\label{sec:mat_meth}

In the following sections, we illustrate the various technical aspects of our work. In particular, Sec.~\ref{subsec:exp_setup} is devoted to the description of the experimental setup; in Sec.~\ref{subsec:tracking} we lay out the pipeline used to reconstruct the motion of the droplets from experimental videos, which involves segmentation and classification, linking, and postprocessing phases; in Sec.~\ref{sec:metrics} we list, describe, and explain the tools employed in analysis of the data.

\subsection{Experimental setup and data acquisition}
\label{subsec:exp_setup}
For the preparation of droplets, stock solutions of 2 mM Oil red O (BioReagent, Sigma-Aldrich) and Sudan black B (certified by the Biological Stain Commission, Sigma Aldrich) in ethyl salicylate (Sigma Aldrich CAS: 118-61-6) were respectively prepared by weighing off appropriate amounts of dyes and adding them to a 50 ml Falcon tube. The dye was solubilized by 10 minutes of sonication at room temperature. The droplet stock solutions were prepared by mixing appropriate amounts of ethyl salicylate, paraffin oil (puriss., CASNumber: 8012-95-1, Sigma-Aldrich), and dye stock solution in 20 ml glass vials to obtain stock solutions of ethyl salicylate with 20wt\% paraffin oil, with 0.001wt\% of either dye. A glass Petri dish with a diameter of $\approx8.6$~cm was placed onto a LED flat panel for controlled illumination. 25 ml of 10g/l aqueous solution of sodium dodecyl sulfate (SDS, Sigma Aldrich, CAS: 151-21-3) was added to the dish using a micropipette obtaining a layer of $\sim$ 4 mm. A tip dipped in the blue droplets solution was inserted for few seconds at the water-air-glass Petri dish interface to prime the surface with ethyl salicylate, preventing initial fission of the droplets. The desired number of 20$\mu$l droplets were placed onto the SDS solution (first the blue droplets and then the red ones) in the Petri dish using a micropipette. The Petri dish was then covered with a 20 cm diameter glass Petri dish. The behavior was filmed using a Logitech CD920 HD Pro Webcam for between 3 and 6 hours. All experiments were performed at lab temperature (19$\pm$1$^{\circ}$C)).

\subsection{Tracking}
\label{subsec:tracking}

Multi-object tracking (MOT) is essential for performing quantitative analyses of video-recorded moving objects. In this work, a semi-automated tracking pipeline has been developed to extract accurate and reliable droplet trajectories from the video footage of the experiments. Aside from the positions of its center, these trajectories contain additional features of each droplet, including its color, eccentricity, and area. The latter can be used to compute the ``radius'' (as perceived from above) of the detected object by approximating it to a circle. Hereafter, we describe each of the three steps involved in the tracking pipeline, namely segmentation and classification, linking, and post-processing.

\subsubsection{Instance segmentation and classification}
\label{subsec:instance_seg}
For each experimental realization of the system, the positions of (the centers of) the droplets, along with information on their shape and color, were extracted \emph{via} instance segmentation and classification performed on every frame of the video recording using the StarDist Python package \cite{schmidt2018, weigert2020, weigert2022}. StarDist employs a neural network that predicts, for each pixel, the probability of being part of an object, the distance of the pixel to the object boundary along several radial directions, and a probability distribution over classes of object categories. The model relies on a slightly modified 3D variant of ResNet \cite{8237584} for the backbone of the network, with two additional $3 \times 3$ convolutional layers with $128$ channels and ReLU activations, each one feeding three output layers; the latter are, respectively, (\emph{i}) the object probability output as single-channel output convolutional layer with sigmoid activation; (\emph{ii}) the polygon distance output as convolutional layer with as many output channels as there are radial directions $n$, with no additional activation function; and (\emph{iii}) the object classification output as a convolutional layer with as many output channels as the number of classes, plus one for the background. For the current application, the StarDist neural network was trained on a diverse set of images extracted from experimental video recordings, covering different lighting conditions and imaging settings. The training set included manually labeled droplet instances to provide ground truth annotations.

\begin{figure}[ht]
    \centering
    \includegraphics[width=\columnwidth]{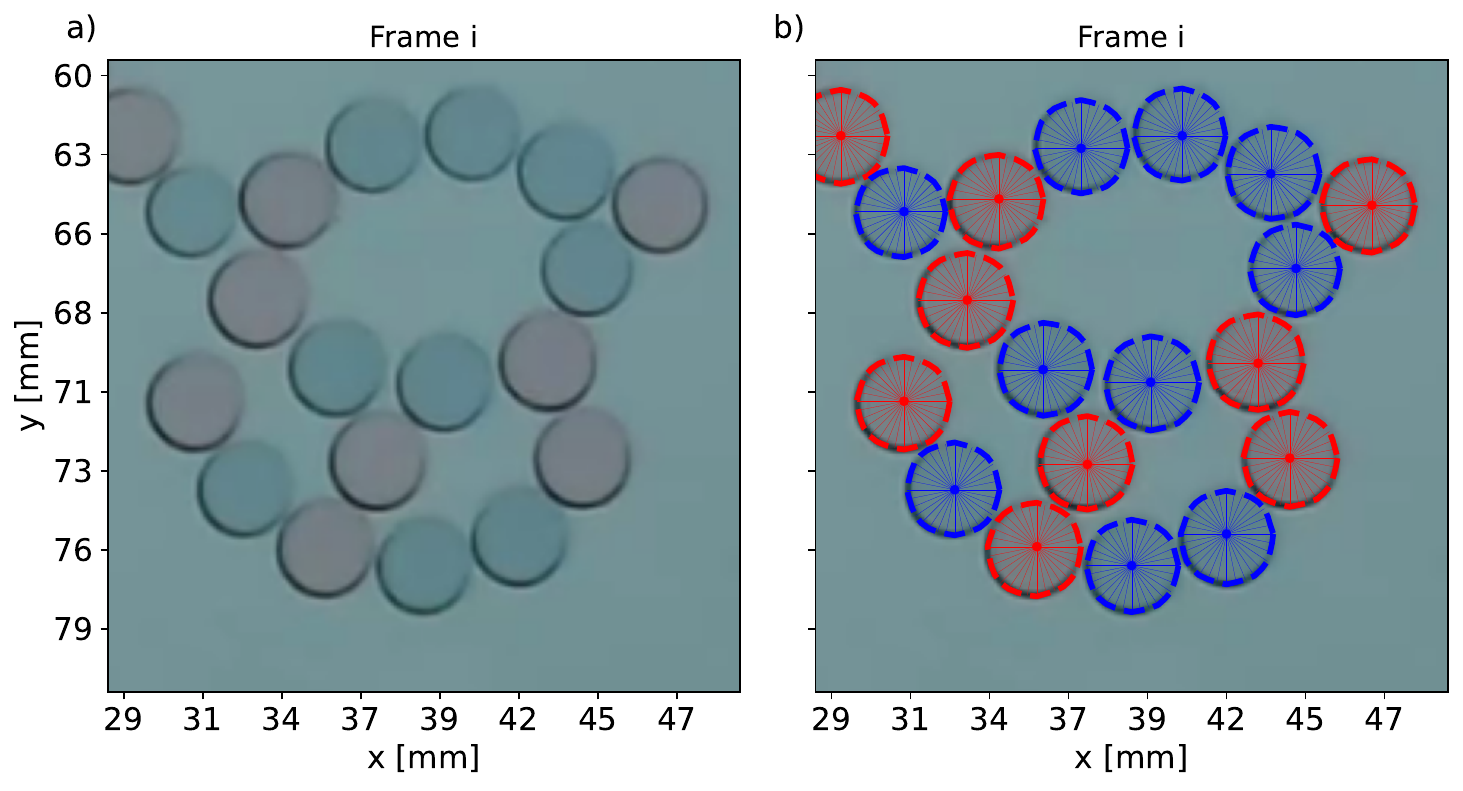}
    \caption{Example of instance segmentation and classification performed with StarDist. (a) Close-up of a portion of an original experimental image. (b) Predicted instance positions and rays overlaid on the image of panel~(a).}
    \label{fig:detection_example}
\end{figure}

In each frame, the features of the detected instances, the latter being either actual droplets or possible spurious objects, were extracted by measuring the properties of the corresponding labeled image. The most important features are the instances' centroid positions, their color, and their area, see Fig.~\ref{fig:detection_example}. Assuming each instance to be circular on the x-y plane, its area allowed us to determine the ``radius'' of the object as perceived from above (hereafter referred to simply as the radius). We note that pixel units were converted into physical ones through image scale calibration. This was achieved by measuring the Petri dish diameter in pixels, comparing it to its known physical diameter, and computing the corresponding scale factor. All relevant instance properties were then rescaled accordingly.

As previously mentioned, at this stage the identified instances comprise actual droplets as well as spurious network detections; in the present case, to filter out most of the latter, it was sufficient to require instances to have a radius greater than $1.25$ mm. Across all experimental realizations, we found that approximately 0.016~\% of frames contained an incorrect number of detected instances after this first filtering step. All encountered errors in a frame involved fewer than $3$ detections of spurious objects (an issue solved in the linking step, \emph{vide infra}) or a single droplet instance missed in the detection (tackled in the post-processing step via interpolation, \emph{vide infra}).

\subsubsection{Linking and trajectory reconstruction}

\begin{figure}[ht]
    \centering
    \includegraphics[width=\columnwidth]{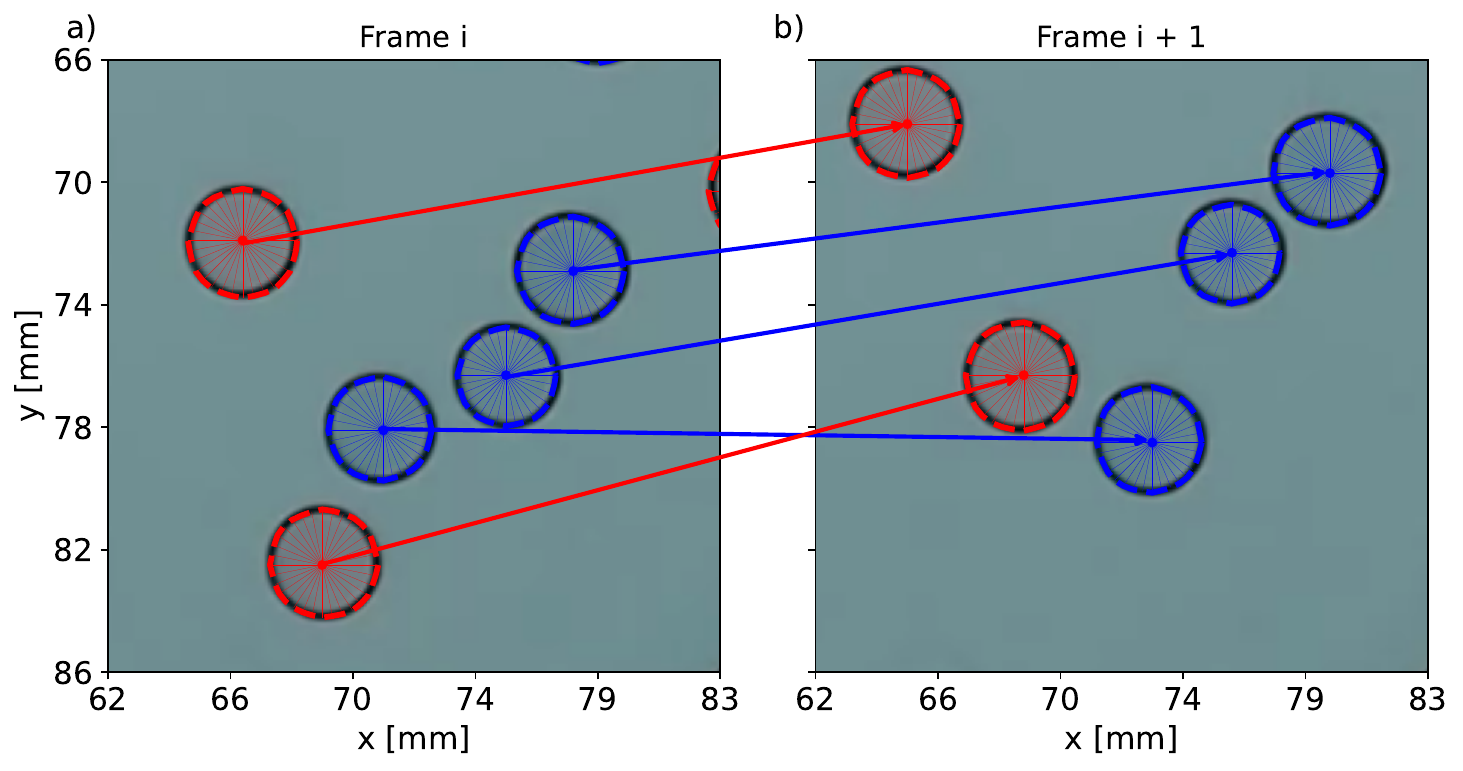}
    \caption{Example of instance linking between consecutive frames of an experimental video. The colors distinguish the droplet class, and the arrows connecting panels (a) and (b) represent the matching of the identities of the droplets between frames $i$ and $i+1$.}
    \label{fig:linking_example}
\end{figure}

The second step of the tracking pipeline consists of linking the detected instances---once filtered as detailed above---across frames while preserving their identities over time, thereby reconstructing the corresponding trajectories. This amounts to determining the most likely assignment of instances between consecutive frames, a process visually depicted in Fig.~\ref{fig:linking_example}. In this work, identity matching was achieved by relying on the initial assumption that the motion of droplets on short timescales---in our case, the ones associated with the frame rate $\Delta t = 1/30$~s of the experimental videos---can be approximated with that of a set of $N$ non-interacting Brownian particles; under such conditions, the probability that the droplets will exhibit a set of displacements $\{\boldsymbol\delta_i\}=(\boldsymbol\delta_1,..,\boldsymbol\delta_N)$ in a time interval $\Delta t$, $P\left(\{\boldsymbol\delta_i\},\Delta t\right)$, reads
\begin{equation}
    P\left(\{\boldsymbol\delta_i\},\Delta t\right) = \left( \frac{1}{4 \pi D \Delta t}\right)^N exp \left( - \sum_{i=1}^N \frac{|\boldsymbol\delta_i|^2}{4 D \Delta t} \right),
    \label{eq:linking}
\end{equation}
where $D$ is the diffusion coefficient. The most likely identity assignment between consecutive frames corresponds to the one that maximizes $P(\{\boldsymbol\delta_i\},\Delta t)$. 
The implementation of this procedure was performed \emph{via} the Trackpy python package \cite{allan_daniel_b_2023_7670439}, which also accounts for the possibility that an instance might be missed in the detection stage for a few frames and then seen again. The resulting data can still contain some spurious, short trajectories of small defects that survived the filtering performed in the segmentation and classification step described in Sec.~\ref{subsec:instance_seg}. This problem was solved by requiring the trajectory of an object to last at least $99~\%$ of the video frames to be attributed to an actual droplet.

\begin{figure}[ht]
    \centering
    \includegraphics[width=\columnwidth]{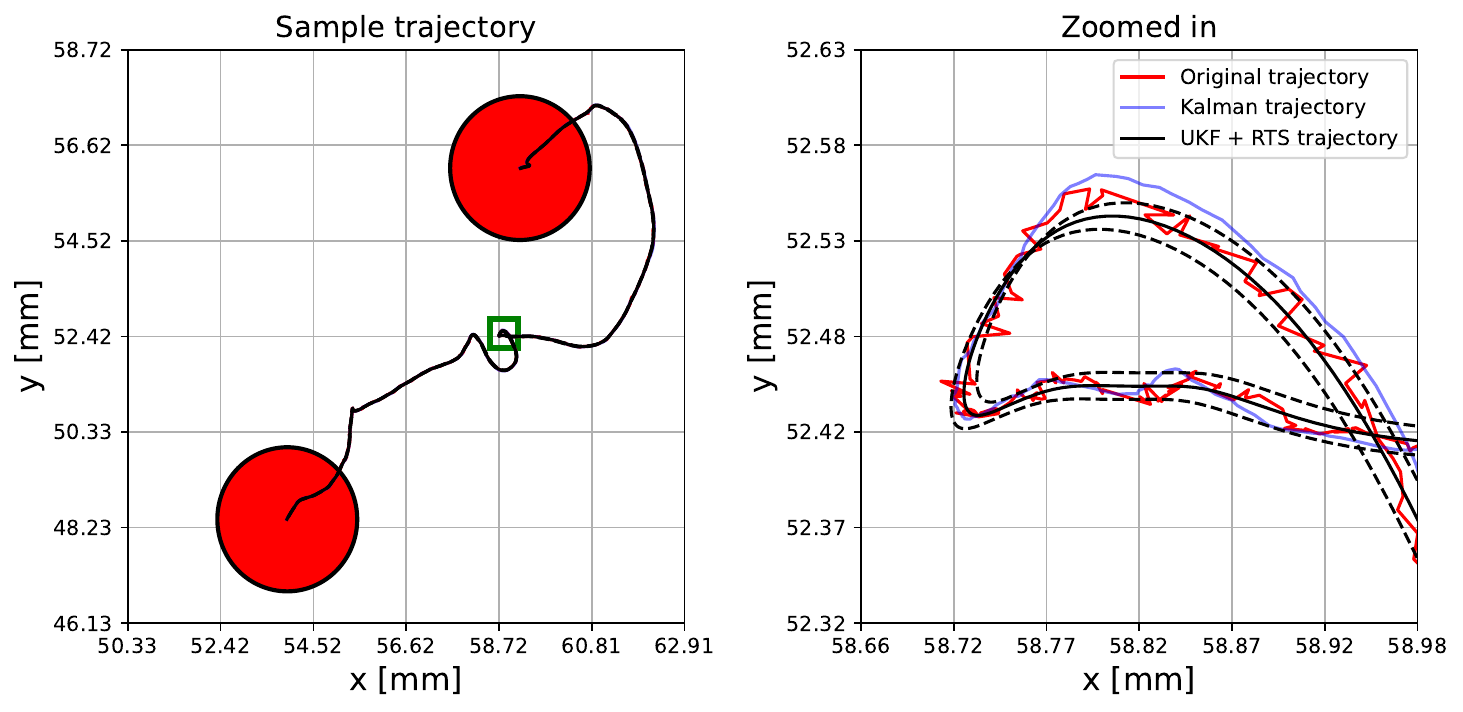}
    \caption{Left: segment of a test droplet trajectory reconstructed by our tracking pipeline. Right: close-up of the region indicated by the green frame in the left panel. We display (\emph{i}) the raw trajectory (red line); (\emph{ii}) the trajectory obtained by applying the unscented Kalman filter to the raw data (blue line); and (\emph{iii}) the trajectory resulting from the unscented Kalman filter combined with the Rauch–Tung–Striebel smoother, with dashed lines indicating the associated confidence interval.}
    \label{fig:kalman_filter_0}
\end{figure}

The end of the linking stage provided us with sub-pixel accurate droplets' trajectories---see  Fig.~\ref{fig:kalman_filter_0} for a graphical example---that were, however, still affected by small-amplitude noise at short timescales. Such noise is not entirely physical and arises also from uncertainties in the StarDist reconstruction of the droplets’ centers, with the latter becoming more relevant once the agents' self-propulsion has largely subsided. The final phase of the tracking pipeline, discussed in the following section, addresses both this issue and that of droplet instances missed by the detection in a few frames of a video, doing so through the application of an interpolation and noise suppression procedure to the trajectories generated by the linking step.

\subsubsection{Post-processing}
\label{subsubsec:post_processing}

The third and final phase of our tracking pipeline is the post-processing of the trajectories obtained at the end of the linking procedure. Here, having already removed previous spurious detections, the motion and features of those droplets that were---for an extremely limited subset of frames, see Sec.~\ref{subsec:instance_seg}---not detected by StarDist were linearly interpolated, and the effect of the network's measurement noise on the trajectories was suppressed using the unscented Kalman filter (UKF) proposed by Julier and Uhlman \cite{10.1117/12.280797}, together with the Rauch-Tung-Striebe (RTS) smoother \cite{doi:10.2514/3.3166}. For the sake of brevity, we will not elaborate on the details of the two algorithms here, referring the interested reader to Ref.~\cite{882463} for their in-depth description.

The UKF is an extension of the standard Kalman filter formulated to perform recursive state estimation in systems with nonlinear dynamic and measurement models, thus overcoming the linearity constraint of the original algorithm \cite{KOUSKOULIS2019782}. In this framework, the system state is represented as a Gaussian random variable (GRV), $\mathbf{x}_k \sim \mathcal{N}(\hat{\mathbf{x}}_k, \mathbf{P}_k)$, where $\hat{\mathbf{x}}_k$ denotes the state estimate and $\mathbf{P}_k$ the error covariance matrix. The state distribution is approximated using a minimal set of deterministically chosen points generated according to Van der Merwe's algorithm \cite{article_Merwe}, which encode the true mean and covariance of the GRV. When transformed by the nonlinear transition model, these points yield posterior mean and covariance estimates that are accurate to third order in a Taylor series expansion \cite{882463}.

Regarding the specifics of our implementation, we estimated the measurement noise covariance matrix $\mathbf{R}$ as
\begin{equation}
    \mathbf{R} = \sigma_r^2 \begin{bmatrix}
        1 & 0  \\
        0 & 1      
        \end{bmatrix}
\end{equation}
where $\sigma_r^2 = 0.01~\text{px}^2$ and considered a simple 2D second-order model with transition matrix $\mathbf{F}$ and measurement matrix $\mathbf{H}$ given by
\begin{equation}
    \mathbf{F} = \begin{bmatrix}
        1 & \Delta t & 0 & 0        \\
        0 & 1        & 0 & 0        \\
        0 & 0        & 1 & \Delta t \\
        0 & 0        & 0 & 1        
        \end{bmatrix}, \quad
    \mathbf{H} = \begin{bmatrix}
        1 & 0 & 0 & 0 \\
        0 & 0 & 1 & 0      
        \end{bmatrix}
\end{equation}
while the process noise covariance matrix $Q$ reads
\begin{equation}
    \scriptstyle{\mathbf{Q} = 
    \begin{bmatrix}
    \sigma_x^2 & \sigma_{x \dot{x}} & 0 & 0 \\
    \sigma_{\dot{x} x} & \sigma_{\dot{x}}^2 & 0 & 0 \\
    0 & 0 & \sigma_y^2 & \sigma_{y \dot{y}} \\
    0 & 0 & \sigma_{\dot{y} y} & \sigma_{\dot{y}}^2 
    \end{bmatrix} = 
    \sigma_a^2 
    \begin{bmatrix} 
    \frac{\Delta t^4}{4} & \frac{\Delta t^3}{2} & 0              & 0 \\ 
    \frac{\Delta t^3}{2}  & \Delta t^2 & 0 & 0 \\ 
    0 & 0 & \frac{\Delta t^4}{4} & \frac{\Delta t^3}{2} \\ 
    0 & 0 & \frac{\Delta t^3}{2} & \Delta t^2 \\ 
    \end{bmatrix}}.
\end{equation}

We underline that, being the UKF a recursive Markovian filter, its state estimate at step $k$ depends only on the prediction at the previous step $k-1$, and so on. To refine the results, we thus also incorporate future measurements into the estimate at step $k$ by relying on an RTS smoother. The latter runs backward over the data and uses the UKF filter prediction along with its covariance matrix to produce smoothed state estimates that optimally account for both past and future observations. 

An example of a trajectory segment produced by this overall workflow in the regime of highest self-propulsion of the droplets is reported in Fig.~\ref{fig:kalman_filter_0}. From the plot, it is possible to appreciate that the protocol suppresses the high-frequency and small-amplitude measurement noise of the neural network, resulting---as long as the activity of the droplets has not completely subsided---in trajectories produced by the tracking pipeline that are characterized by high resolution and accuracy. When the activity of the agents has largely vanished and these have become essentially steady, on the other hand, the network's inaccuracies in the detection of the droplet centers can impact the results of the trajectory reconstruction beyond the 30fps timescales of the experimental videos; this makes it difficult, at such a framerate, to distinguish real motion from noise. To address this issue, the output trajectories of our pipeline were finally subsampled from the original 30 fps recording to 10 fps. By reducing the temporal resolution, we effectively filter out small-timescale processes that cannot be reliably resolved in the phase in which droplets are largely immobile, ensuring that all the subsequent analyses are based on dynamical data that are robust and physically meaningful throughout the totality of the experiment.

\subsection{Trajectory windowing and analyzed metrics}
\label{sec:metrics}

The natural statistical mechanics approach for investigating a non-equilibrium system involves performing ensemble averages of physical observables over multiple realizations of its time evolution to gain insight into its statistical behavior \cite{Zwanzig2001}. This framework can be complemented if the system reaches a time-translationally invariant state---as occurs, for instance, under stationary non-equilibrium conditions; in such a case, if the system is ergodic with respect to such stationary distribution, ensemble averages over the latter can be replaced by time averages along a single, long trajectory, discarding the initial relaxation transient.

In the case under examination, the droplets---essentially hydrodynamic entities---were mapped onto point-like particles, and the pipeline described in Sec.~\ref{subsec:tracking} provided highly accurate trajectories of the agents extracted from the experimental videos. While such a mapping renders the system amenable to a statistical-mechanics-based analysis, the limited number of realizations available, together with the small number of droplets in the experimental setup considered, precludes a proper ensemble-averaged approach. Moreover, the progressive decay of self-propulsion prevents the system from reaching a stationary non-equilibrium state, except for the dynamically trivial inactive regime in which the agents have become essentially immobile; consequently, time-averaging along a single, long trajectory is not a feasible strategy for characterizing the system's statistical behavior.

Nevertheless, the macroscopic timescales over which the activity vanishes (on the order of three hours) are much larger than the characteristic ones of the droplets' dynamics we aim to investigate. To resolve the time dependence of the system’s statistical properties along its evolution, we hence rely on a \emph{window-based analysis}. Specifically, we introduce time windows of duration $T_w=600$~s and assume that, on such a timescale, the self-propulsion properties of the system do not change significantly---a condition not strictly valid in the initial regime of high activity---and produce only a somewhat “adiabatic” modification of the droplets’ short-time dynamics. Furthermore, it is assumed that, for approximately constant self-propulsion, the system relaxes rapidly toward a non-equilibrium steady state. Under these strong assumptions of overall ``quasi-stationary non-equilibrium conditions''---holding on timescales short compared to those of self-propulsion decay---we replace ensemble averages over independent realizations of the system characterized by comparable self-propulsion with time averages within each window, sliding the beginning of the latter along the whole trajectory by a stride of $10$ s. Within this window-based framework, a series of time-dependent metrics is then employed to extract insight on the system's behavior as its activity wanes, including structural and dynamic observables. Such metrics are, respectively: (\emph{i}) the average speed of the droplets; (\emph{ii}) their velocity polarizations; (\emph{iii}) their hexatic order parameters; (\emph{iv}) their time- and species-averaged mean squared displacements  (TSAMSD); (\emph{v}) their velocity autocorrelation functions (VACF); (\emph{vi}) their turning angle distributions; and (\emph{vii}) the dimer distributions introduced by Tanaka and coworkers \cite{doi:10.7566/JPSJ.86.101004}.

The droplets' average speed is employed in this work as a proxy to quantify their level of self-propulsion over the course of the experiment. The average speed of an agent is defined as
\begin{equation}
    \label{eq:speed}
    V_{\mathcal{S}}(t_w)=\langle \langle||\mathbf{v}_k(t)||\rangle_{k\in\mathcal{S}}\rangle_{t\in{T_w}},
\end{equation}
where $\mathbf{v}_k(t)=[\mathbf{x}_k(t+\Delta t)-\mathbf{x}_k(t)]/{\Delta t}$ is the velocity of droplet $k$ at time $t$ as extracted from the reconstructed trajectory, $\Delta t$ is 0.1 s as for the subsampling described in \ref{subsubsec:post_processing}, while $\langle \cdot \rangle_{t\in T_w}$ indicates, in the following, the time average of the bracketed observable over a window of length $T_w$ centered around the running time $t_w$. Given the different behavior observed for the droplets depending on their color, see Fig~\ref{fig:fig1}, we further consider a global speed parameter associated with the whole system as well as an independent one for each species. In this respect, the symbol $\langle \cdot \rangle_{k \in \mathcal{S}}$ in Eq.~\ref{eq:speed} indicates---and does so in the remainder of this work---an average of the bracketed observable performed over all droplets ($\mathcal{S}=\mathcal{A}$) or restricted to the blue and red agents ($\mathcal{S}=\mathcal{B}$ and $\mathcal{R}$, respectively).

The degree of collective orientational alignment in the motion of the droplets is here analyzed in terms of their velocity polarization $\Phi$, which acts as an order parameter in the Vicsek model \cite{PhysRevLett.75.1226}. Accounting for the possible groupings $\mathcal{S}=\mathcal{A},\mathcal{B},\mathcal{R}$ of species identities, the set of velocity polarizations considered in this work reads
\begin{eqnarray}
    \Phi_\mathcal{S}(t_w) &&= \left\langle \left\lVert \left\langle \frac{\mathbf{v}_k(t)}{\lVert \mathbf{v}_k(t) \rVert} \right\rangle_{k \in \mathcal{S}} \right\rVert \right\rangle_{t\in T_w} \nonumber \\
    &&= \langle \lVert \langle \hat{\mathbf{e}}_k(t) \rangle_{k \in \mathcal{S}} \rVert \rangle_{t\in T_w},
    \label{eq:order}
\end{eqnarray}
where $\hat{\mathbf{e}}_k(t)$ is the versor of the velocity of droplet $k$ (or its \emph{orientation} in what follows) at time $t$. We note that in the case of a perfect alignment of the velocities of all droplets within the group $\mathcal{S}$, one obtains that the corresponding polarization is $\Phi_\mathcal{S}=1$; on the other hand, random, uncorrelated orientations result in
\begin{equation}
    \Phi_\mathcal{S} = \left\langle \frac{1}{N_\mathcal{S}} \sqrt{\sum_{k \in \mathcal{S}} |\hat{\mathbf{e}}_k|^2 + \cancel{\sum_{k\neq j \in \mathcal{S}} \hat{\mathbf{e}}_k\cdot \hat{\mathbf{e}}_j}} \right\rangle_{t\in T_w} =\frac{1}{\sqrt{N_\mathcal{S}}},
    \label{eq:order_uncorrelated}
\end{equation}
and since, in the present case, the population consists of $N_{\mathcal{B}} = 25$ blue droplets and $N_{\mathcal{R}} = 25$ red ones, their disordered dynamics is associated with a set of baseline velocity polarizations of $\Phi_\mathcal{\mathcal{B},\mathcal{R}}=0.2$ and $\Phi_\mathcal{\mathcal{A}} \approx 0.14$.

Albeit sensitive to the presence of collectiveness in the droplet's motion, the velocity polarization is instead possibly blind to what are eventual regularities in their \emph{spatial} arrangement. To capture the latter, we here rely on (a slightly modified version of) the hexatic order parameter $\Psi_6$ measuring how closely, on average, the local environment around a droplet resembles perfect hexagonal/hexatic symmetry \cite{jaster2004hexatic}. Taking into account all possible groupings of droplets, the resulting set of hexatic order parameters is here defined as
\begin{equation}
    \Psi_{6,\mathcal{S}}(t_w) = \text{Re}\left[\left\langle \left\langle \frac{1}{6} \sum_{j\in\mathcal{S} = 1}^{n_k(t)} e^{i6  [\eta_{kj}(t) - \eta_{k1}(t)]} \right\rangle_{k\in \mathcal{S}}\right\rangle_{t\in T_w}\right],
    \label{eq:hex_order}
\end{equation}
where $n_k$ is the number of $\mathcal{S}$-neighbors of droplet $k\in{\mathcal{S}}$, and $\eta_{kj}$ is the angle that the displacement vector $\mathbf{r}_{kj}$ connecting the position of droplet $k$ with that of its $j$-th neighbor forms with the $x$ axis. We consider a droplet $j$ to be in the neighborhood of $k$ if $|\mathbf{r}_{kj}| < 1.5 d_S$, where $d_S$ is the mean diameter of the droplets of species $S$ at the specific frame considered. Again, time averages in Eq.~\ref{eq:hex_order} are performed over windows of length $T_w$ centered around the running time $t_w$, and the metric is computed separately for blue and red droplets ($\mathcal{S}=\mathcal{B},\mathcal{R}$), as well as among all droplets ($\mathcal{S}=\mathcal{A}$). Critically, $\Psi_6$ also captures the presence of \emph{partial} hexatic order in the system---note in Eq.~\ref{eq:hex_order} the multiplicative factor of $1/6$ rather than $1/n_k$---in that $q$ neighbors located on the edges of a hexagon centered around droplet $k$ result for it in $\Psi_{6,k} = q/6$. The recalibration of the neighbors' angles with respect to the first of them taken as reference was introduced to render the metric rotationally invariant.

The next step in the analysis involves investigating the translational and rotational diffusive properties of the droplets and how these are influenced by the decay of self-propulsion. As for translational ones, we calculate the time- and species-averaged mean squared displacements (TSAMSD) $\delta_{\mathcal{S}}^2$ and velocity autocorrelation functions (VACF) $C_{\mathcal{S}}$ of the two populations $\mathcal{S}=\mathcal{B},\mathcal{R}$ as a function of the running window time $t_w$ and the lag time $\tau$. These TSAMSDs and VACFs read, respectively
\begin{eqnarray}
    \delta_{\mathcal{S}}^2(\tau,t_w) = \langle \langle (\mathbf{r}_k(t+\tau) - \mathbf{r}_k(t))^2 \rangle_{t\in T_w} \rangle_{k \in \mathcal{S}}\ ,
    \label{eq:taemsd} \\
    C_{\mathcal{S}}(\tau,t_w) = \frac{1}{\sigma_{ \mathcal{S}}^2(t_w)} \big\langle \left\langle \delta\mathbf{v}_k(t) \cdot \delta\mathbf{v}_k(t+\tau) \right \rangle_{t \in T_w} \big\rangle_{k \in \mathcal{S}}\ .
    \label{eq:VACF}    
\end{eqnarray}
In Eqs.~\ref{eq:taemsd} and~\ref{eq:VACF}, $t_w$ is again the running time located at the center of the window of length $T_w$ over which the time average $\langle \cdot \rangle_{t\in T_w}$ is performed. Furthermore, $\delta\mathbf{v}_k(t)=\mathbf{v}_k(t) - \langle \mathbf{v}_k(t) \rangle_{t \in T_w}$, while $\sigma_{\mathcal{S}}^2(t_w)$ is the variance of the velocity of species $\mathcal{S}$ within the time window.

The rotational diffusive properties of the system are instead analyzed \emph{via} the droplet species' distributions of turning angle $\Delta\theta$, the latter being the angle between two consecutive displacements of a droplet---note that, for each agent, the instantaneous direction of its motion is defined by the orientation $\hat{\mathbf{e}}_k(t)=(\hat{e}_{k,x}(t),\hat{e}_{k,y}(t))$ of its velocity defined in Eq.~\ref{eq:order}. Starting from the droplet trajectories, the distribution $P_{\mathcal{S}}(\Delta \theta,t_w)$ of species $\mathcal{S}$ at running time $t_w$ is calculated as
\begin{eqnarray}
        &&P_{\mathcal{S}}(\Delta \theta,t_w) = \left\langle \left\langle  \delta(\theta_k(t+\Delta t) - \theta_k(t) - \Delta \theta)\right\rangle_{t \in T_w} \right\rangle_{k \in \mathcal{S}}\ , \nonumber \\
        &&\theta_k(t) = \text{atan2}\left[{\frac{\hat{e}_{k,y}(t)}{\hat{e}_{k,x}(t)}}\right],
    \label{eq:turn_angl}
\end{eqnarray}
where $\text{atan2}$ is the extension of the conventional $\arctan$ function with image in $[-\pi;\pi]$, the difference $\theta_k(t+\Delta t) - \theta_k(t)$ has to be interpreted mod $2\pi$ (so that $\Delta \theta \in [-\pi,\pi]$), and $\Delta t=0.1$~s as for the sub-sampling described in \ref{subsubsec:post_processing}.

Finally, the \emph{global} structural arrangement of the system was analyzed by computing the average 2D distribution of droplets around a \emph{dimer} \cite{doi:10.7566/JPSJ.86.101004}, varying the composition of the latter to include blue-blue, red-red, as well as inter-species blue-red and red-blue pairs. Specifically, given the configuration of the system at time $t$, the distribution $D_{\mathcal{S}\mathcal{S'}} (t)$ of droplets around $\mathcal{S}$-$\mathcal{S}'$ dimers, with $\mathcal{S}, \mathcal{S'} \in (\mathcal{B},\mathcal{R})$, is computed by
individually running over all the $\alpha=1,...,N_{\mathcal{S}}$ droplets in species $\mathcal S$, and for each of them applying a roto-translation of the whole system that brings the $\alpha$ droplet in the origin and its nearest neighbor droplet from species $\mathcal{S}'$ (the other member of the dimer) on the positive $x$ axis; to construct $D_{\mathcal{S}\mathcal{S'}} (t)$, a 2D histogram of the droplet positions is subsequently generated that, however, only takes into account the $\mathcal{S}'$ species. The window-based version of this metric is then nothing but a time average of its instantaneous counterpart performed over the extent $T_w$ of the window, namely
\begin{equation}
    D_{\mathcal{S}\mathcal{S'}}(t_w) = \langle D_{\mathcal{S}\mathcal{S'}}(t) \rangle_{t \in T_w}.
    \label{eq:dimer}
\end{equation}

\section{Results}
\label{sec:res_disc}

Although characterized by a relatively plain chemical composition, the active droplet system analyzed in this work exhibits---as already highlighted by Tanaka and coworkers \cite{doi:10.7566/JPSJ.86.101004} in the single-component case---an extremely complex, life-like behavior during its temporal evolution, unfolding as the self-propulsion gradually decays. As highlighted in Fig.~\ref{fig:fig1}, a gradual increase in collective order is observed from a purely structural standpoint along the experiment; specifically, the system starts from an initial state of almost random and uniform spatial arrangement of the droplets within the Petri dish, then passes through the generation of transient structures for intermediate times, and finally reaches a phase of maximal, closely packed, and quasi-crystalline hexagonal order. Critically, this overall growth in structural regularity is accompanied by, and intertwined with, an evolution of the system's dynamical properties: the disordered behavior distinctive of the initial, high-activity regime progressively gives way to an increasingly coordinated motion of the agents, lastly reaching an essentially static phase when their propulsion has subsided. Furthermore, the blue and red droplets manifest these structural and dynamic features differently due to their uneven activity levels---the one of the red agents being significantly higher during the early steps of the experiment---and distinct rates of activity decay; starting from the homogeneous distribution of the two species in the dish observed in the early phase of the system's evolution, a gradual demixing transition occurs, eventually resulting, in the steady regime, in all blue droplets being organized in a cluster surrounded and caged by a chain of red ones, see Fig.~\ref{fig:fig1}.

This sequence of steps, followed by the system over time---uncoordinated motion, generation of transient structures, and finally steady, demixed, and quasi-hexagonal order---appears to be qualitatively preserved across different experimental realizations, see Fig.~S1 of Supporting Information. It is hence natural to begin our analysis looking for simple collective variables in terms of which we can characterize the ``life-to-death cycle'' of the system and adequately capture the interplay of its structural and dynamical features.

\subsection{Collective variables analysis and  identification of the evolutionary stages}
\label{sec:coll_var}

Based on the preceding observations, the trajectories of the droplet centers extracted by our tracking pipeline were first inspected in terms of two sets of collective variables, namely the velocity polarizations $\Phi_{\mathcal{S}}(t_w)$ and hexatic order parameters $\Psi_{6,\mathcal{S}}(t_w)$ respectively reported in Eq.~\ref{eq:order} and~\ref{eq:hex_order} of Sec.~\ref{sec:metrics}. The analysis of these quantities for both the system as a whole ($\mathcal{S}=\mathcal{A}$) and separately by species (red/blue, $\mathcal{S}=\mathcal{B},\mathcal{R}$), is further flanked by the study of the droplets' average speeds $V_{\mathcal{S}}(t_w)$ defined in Eq.~\ref{eq:speed}, the latter enabling the interpretation of the orientational collective behavior of the agents' motion---as captured by the polarizations $\Phi_{\mathcal{S}}$---in the light of the amount of their self propulsion. We remind, in passing, that all time-dependent quantities under examination are defined as the time average of the corresponding instantaneous value, performed in a sliding window of duration $T_w=600$~s centered around the running time $t_w$.

The aforementioned metrics provide complementary information on the collective properties of the system over the course of its evolution. Specifically, the hexatic order parameters $\Psi_{6,\mathcal{S}}$ quantify the degree of \emph{spatial} hexagonal order in the droplets' arrangement, and are thus adequate to signal the emergence of structural regularity in the system. On the other hand, the $\Psi_{6,\mathcal{S}}$ are insensitive to any coordination in the droplets' motion; this aspect is instead effectively captured by the velocity polarizations $\Phi_{\mathcal{S}}$, which are commonly employed to quantify the degree of collective orientational order in the dynamics of active agents (such as flocks of birds, swarms of insects, and schools of fishes \cite{doi:10.1073/pnas.1005766107, biomimetics9110660, doi:10.1073/pnas.2406293121}) while remaining blind to possible symmetries in their spatial configuration. Furthermore, the values attained by the velocity polarizations reflect the coherence of the droplets' motion \emph{irrespective of its intensity}: hence, it is advisable to interpret the $\Phi_{\mathcal{S}}$ in the light of the average droplets' speeds $V_{\mathcal{S}}$.
In conclusion, remodulations over time in the collective structural and dynamical properties of the system can thus be adequately accounted for \emph{via} the simultaneous investigation of the collective variables $\Psi_{6,\mathcal{S}}$ and $\Phi_{\mathcal{S}}$, plus the average speeds $V_{\mathcal{S}}$.

\begin{figure}[t]
    \centering 
    \includegraphics[width=\columnwidth]{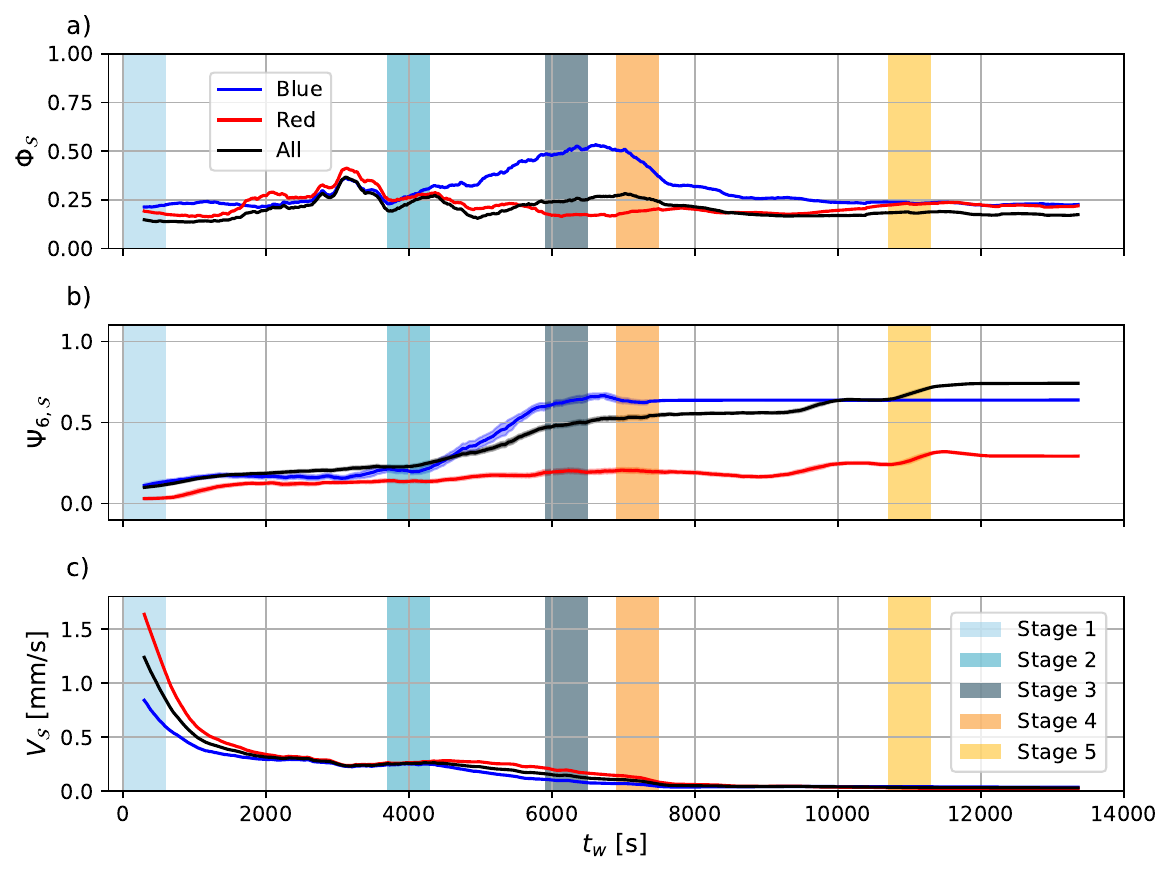}
    \caption{Temporal evolution of (a) the velocity polarization $\Phi_{\mathcal{S}}$, see Eq.~\ref{eq:order} (b) the hexatic order parameter $\Psi_{6,\mathcal{S}}$, see Eq.~\ref{eq:hex_order}, and (c) the average speed $V_{\mathcal{S}}$, see Eq.~\ref{eq:speed}, of the blue droplet population (blue line), the red droplet population (red line), and overall droplet population (black line) as a function of the window running time $t_w$. Colored vertical bars indicate the onset of each evolutionary stage of the system.}
    \label{fig:order}
\end{figure}

The evolution of these quantities over the course of a representative experiment is shown in Fig.~\ref{fig:order}, with corresponding data from two additional realizations being provided in the Supporting Information. Critically, a qualitatively consistent behavior is observed for the temporal trend followed by the observables across the three experiments, exhibiting distinct, recognizable phases. These phases are not necessarily stationary---within each of them, a given collective variable may increase, plateau, or decrease---yet they remain clearly distinguishable from one another. Hence, it is conceptually tempting and practically convenient to partition the system's history into five ``evolutionary stages'': these are highlighted in Fig.~\ref{fig:order}, where vertical colored bars indicate the characteristic window running times $t_w$ marking the beginning of each stage. For times $t_w$ located in the middle of each bar in Fig.~\ref{fig:order}, specific observables representative of the early moments of each stage will be showcased in the following. 

\paragraph{\textbf{Stage I} \textemdash } The first stage (hereafter referred to as I) identified through an inspection of the temporal behavior of the set of $\Phi_{\mathcal{S}}$, $\Psi_{6,\mathcal{S}}$ and $V_{\mathcal{S}}$ is the one displaying the highest self-propulsion of the droplets, and is characterized by an overall absence of collective and persistent dynamic or structural order in the system, both at the global level and when analyzed by individual species (see Fig.~\ref{fig:order}). The high mobility of the agents in this stage can be evinced by their average speeds $V_{\mathcal{S}}$ presented in Fig.~\ref{fig:order}c, which, although decreasing as the stage unfolds, attain the maximum values observed throughout the whole experiment. As for the lack of collective order, this is, from a dynamic standpoint, witnessed by velocity polarizations $\Phi_{\mathcal{S}}$ $\mathcal ({S}=\mathcal{A},\mathcal{B},\mathcal{R}$) on average oscillating around $0.1-0.2$, except for a mild ``bump'' that appears towards the end of the stage; according to Eq.~\ref{eq:order_uncorrelated}, this is indicative of an uncorrelated motion of the droplets. The absence of collective structural regularity in their spatial arrangement can instead be inferred by almost constant hexatic order parameters $\Psi_{6,\mathcal{S}}$ along the stage and lying below $2/6$, with droplets hence having, on average and irrespective of their species, fewer than two nearest neighbors satisfying hexagonal symmetry, see Sec.~\ref{sec:metrics}.

In terms of phenomenological behavior of the droplets, stage I extends from the initial, uncoordinated phase of the system's spatial and dynamic organization to the one in which transient structures progressively emerge, where, in both cases, the two species are found to be approximately homogeneously mixed, see Fig.~\ref{fig:fig1}a-b. Based solely on the analysis of velocity polarizations and hexatic order parameters during this stage, no extremely marked differences between the two species are apparent, suggesting that red and blue agents play an essentially equivalent role in both the uncoordinated and transient-structure regimes as far as these metrics reveal. However, an inspection of the droplets' average speeds $V_{\mathcal{S}}$ reported in Fig.~\ref{fig:order}c indicates that the activity of the red species is significantly higher than that of its blue counterpart at the onset of the experiment and in the uncoordinated phase. $V_{\mathcal{B}}$ and $V_{\mathcal{R}}$, on the other hand, decrease and become gradually comparable throughout stage I, plateauing towards its end to a common value; interestingly and consistently with what observed \emph{via} other metrics (see below), this suggest that the two droplet species display similar dynamical properties in the series of transient structures that appear for intermediate activities.

\paragraph{\textbf{Stage II} \textemdash } At longer timescales and as the self-propulsion of the agents further weakens, the system transitions from this initial, overall disordered, and mixed phase into what we refer to as the second stage (II) of its evolution, see Fig.~\ref{fig:order}. The defining feature of this stage lies, compared to what was observed in I, in the emergence of a clear discrimination in the collective properties of the two species as quantified by their respective $\Phi_{\mathcal{S}}$ and $\Psi_{6,\mathcal{S}}$. In particular, during stage II, an ordering in the dynamics of the blue droplets progressively develops over time, with their velocity polarization rising from the uncorrelated baseline of $\Phi_{\mathcal{B}}=0.2$ to approximately $0.5$. This organization of the dynamics is accompanied by a gradual increase in the symmetry of the blue species' spatial arrangement; this is indicated by the hexatic order parameter $\Psi_{6,\mathcal{B}}$ reaching $\sim 0.6$ by the end of stage II, corresponding to blue droplets being, on average, surrounded by four neighbors of the same color located on the edges of a hexagon. The red species, in contrast, persists in its disordered motion along II, as reflected by values of $\Phi_{\mathcal{R}}$ and $\Psi_{6,\mathcal{R}}$ similar to the ones observed during stage I of the system's evolution.

\begin{figure*}[t]
    \centering
    \includegraphics[width=\textwidth]{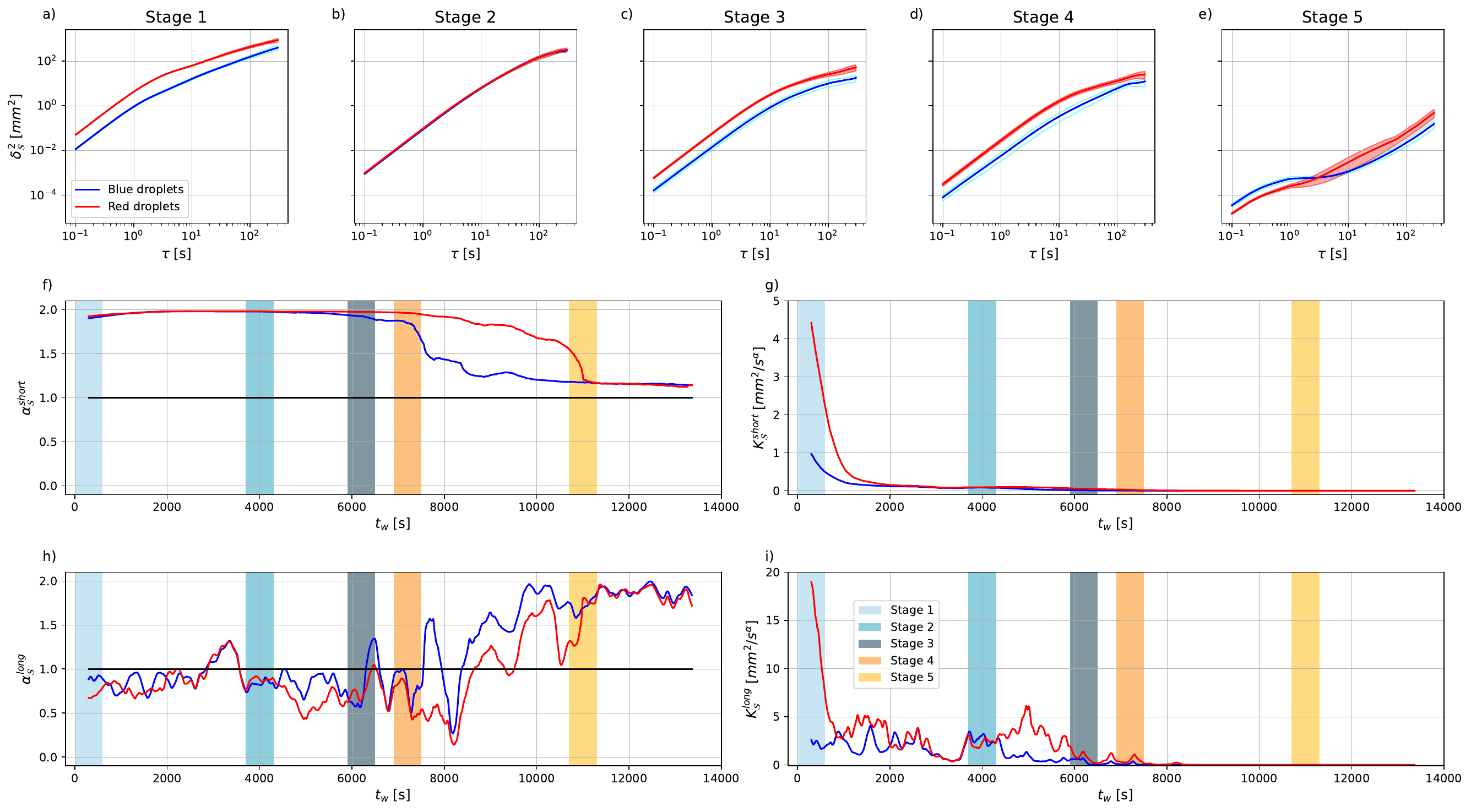}
    \caption{Results for the time- and species-averaged mean squared displacements (TSAMSD) of the two droplet populations, see Eq.~\ref{eq:taemsd}. (a-e) Progression from left to right illustrates the dependence of the family of $\delta_{\mathcal{S}}^2(\tau,t_w)$ TSAMSD curves on the lag time $\tau$ for the blue (blue line) and red (red line) species at the onset of each evolutionary stage of the system---that is, for window times $t_w$ located in the middle of each of the five colored vertical bars in the subsequent plots. (f-g) Temporal evolution of the short time-lag scaling exponents $\alpha_{\mathcal{S}}^{short}(t_w)$ and generalized diffusion coefficients $K_{\mathcal{S}}^{short}(t_w)$ of the two species, resulting from a power law fit, see Eq.~\ref{eq:power_law_fit}, of the family of $\delta_{\mathcal{S}}^2(\tau,t_w)$ profiles of the two species performed in the region $\tau \in [0.1, 1]$~s. Colored vertical bars indicate the onset of each evolutionary stage. (h-i) Temporal evolution of the long-time lag scaling exponents $\alpha_{\mathcal{S}}^{long}(t_w)$ and generalized diffusion coefficients $K_{\mathcal{S}}^{long}(t_w)$ of the two species, obtained from a power law fit of the blue and red droplets TSAMSD performed in the region $\tau \in [50, 300]$~s.}
    \label{fig:TAEMSD_stages}
\end{figure*}

As previously discussed, the late offset of stage I features, as does the beginning of stage II, red and blue agents with similar and quasi-stationary activities that participate in the generation of a series of transient structures displaying an approximately homogeneous species composition, see Fig.~\ref{fig:fig1}b. With the resumption of the decline of self-propulsion during II---evidenced by the further reduction in the droplets' average speeds throughout this stage, faster for the blue component---these structures gradually give way to a demixing transition of the two species, culminating in the formation of a cluster of blue droplets with hexagonal symmetry that accounts for the the steady rise of $\Psi_{6,\mathcal{B}}$ observed in Fig.~\ref{fig:order}. During its growth, the cluster gets increasingly surrounded by the expelled red agents, whose disordered spatial arrangement is witnessed by the quasi-constancy of $\Psi_{6,\mathcal{R}}$ across stages I and II. As for the system's dynamics, the \emph{individual} motion of the blue droplets \emph{within} the cluster becomes, along stage II, more and more subdued over time and as their activity continues to decline; on the other hand, the cluster \emph{as a whole} gradually acquires a \emph{collective} orientational order in its motion, driven by the asymmetric ``kicks'' it receives from the surrounding and still relatively active red component---this being not only structurally, but also dynamically disordered, where we note that $\Phi_{\mathcal{R}} \approx 0.2$ in II. This interplay between the two species underlies the increase in the velocity polarization of the blue droplets observed in Fig.~\ref{fig:order} throughout this stage.

\paragraph{\textbf{Stage III} \textemdash } The timescales at which the previously discussed steady growth in the hexatic order parameter and velocity polarization of the blue species halts mark the onset of the third stage of the system’s evolution. During stage III, all the collective variables $\Phi_{\mathcal{B}/\mathcal{R}}$ and $\Psi_{6,\mathcal{B}/\mathcal{R}}$ remain approximately constant, see Fig.~\ref{fig:order}. The cluster of blue droplets is fully formed, will persist for the remainder of the experiment, and its constituents have lost most, if not all, of their spontaneous self-propulsion. As in the late phase of stage II, the residual collective motion of blue droplets, highlighted again by a $\Phi_{\mathcal{B}}$ lying around $0.5$, arises primarily from the random kicks that are imparted to the cluster by the surrounding red droplets, which remain both relatively active---though exhibiting a further decline in self-propulsion, evidenced by a decrease in $V_{\mathcal{R}}$ along the stage---and dynamically/structurally disordered.

\paragraph{\textbf{Stage IV} \textemdash } As the red species' self-propulsion ultimately vanishes, see Fig.~\ref{fig:order}c, the system takes its steps towards the terminal, ``death'' phase. In particular, stage IV is marked by a decline in the blue droplets’ velocity polarization, which steadily decays over time to the baseline value of $0.2$ (see Fig.~\ref{fig:order}a) signaling a return to uncorrelated motion---though with subtle differences to be discussed below. The decay of $\Phi_{\mathcal{B}}$ occurs at almost constant hexatic order parameter $\Psi_{6,\mathcal{B}}$ and red species’ velocity polarization $\Phi_{\mathcal{R}}$, and is instead accompanied by a mild increase in $\Psi_{6,\mathcal{R}}$. A progressive extinction of the collective dynamics of the stable blue cluster is hence observed during IV; critically, this is driven by the red droplets losing their residual self-propulsion and contextually adopting a more symmetric arrangement around the cluster---note that $\Psi_{6,\mathcal{R}}\approx 0.3$ at the end stage IV, with each red droplet having on average two nearest neighbors located at its sides. As a result, the persistent and inhomogeneous ``kicks'' provided by the red species become increasingly weaker and more isotropic, eventually ceasing to drive the global displacement of the blue cluster.

\paragraph{\textbf{Stage V} \textemdash} When the self-propulsion of the red droplets has fully subsided, the system is found in its final phase, indicated in Fig.~\ref{fig:order} as stage five (V) of the evolution. Here, the hexatic order parameters and velocity polarizations of the two species remain largely constant over time, retaining the values they reached at the end of stage IV. In particular, both $\Phi_{\mathcal{B}}$ and $\Phi_{\mathcal{R}}$ settle at the $0.2$ baseline characteristic of an uncorrelated motion, see Fig.~\ref{fig:order}; unlike in the high-activity regime, however, this value now arises from droplets that are almost entirely immobile (see the agents' average speeds reported in Fig.~\ref{fig:order}c), so that the observed disorder in the orientation of their velocities is primarily dictated by residual noise of small amplitude in the detection of their positions. From a structural standpoint, stage V corresponds to a final, largely steady and quasi-crystalline configuration of the system, see Fig.~\ref{fig:fig1}e, in which the hexagonally ordered cluster of blue droplets is caged by a ring of red ones.

\subsection{Time- and species-averaged mean squared displacements \& Velocity autocorrelation functions}

In Sec.~\ref{sec:coll_var} we highlighted how it is possible---also for practical convenience---to partition the global evolution of the system into five distinct stages based on the phenomenological behavior followed by the set of collective variables $\Phi_{\mathcal{S}}$, $\Psi_{6,\mathcal{S}}$, and $V_{\mathcal{S}}$ over time, stages that consistently emerge across all three experimental runs with only minor variations (see Supporting Information). Building on this partition, the remainder of this work will be devoted to analyzing a set of complementary observables that capture different aspects of the system’s dynamical and structural properties, paying particular attention to the influence that the decrease in self-propulsion has on these metrics. We begin this discussion focusing on the time- and species-averaged mean squared displacements (TSAMSDs, see Eq.~\ref{eq:taemsd}) and velocity autocorrelation functions (VACFs, see Eq.~\ref{eq:VACF}) of the agents, which together provide insight into the droplets’ translational diffusive features and the persistence of their motion.

To characterize the dynamics of the agents and its dependence on the gradual loss of activity, we thus first evaluated the TSAMSDs of the droplets $\delta_{\mathcal{S}}^2(\tau,t_w)= \langle \langle (\mathbf{r}_k(t+\tau) - \mathbf{r}_k(t))^2 \rangle_{t\in T_w} \rangle_{k \in \mathcal{S}}$. We remind that $\mathcal{S}=\mathcal{B},\mathcal{R}$ marks the species identity, while $t_w$ is the running time located at the center of the window over which the time average $\langle\cdot\rangle_{t\in T_w}$ is accumulated.

MSDs constitute a fundamental tool for investigating self-propelled systems, allowing one to identify the presence of distinct dynamical regimes---such as ballistic motion, normal/anomalous diffusion, or arrest---that may emerge across different timescales \cite{howse2007self, breoni2020active}. A well-known example of this is provided by non-interacting active Brownian particles: here, theoretical predictions (neglecting short-time diffusion) \cite{howse2007self} yield $\delta^2(\tau)\propto \tau^2$ for time lags $\tau \ll \tau_r = 1/D_r$ where $D_r$ is the particles' rotational diffusion coefficient, corresponding to ballistic motion, while a diffusive regime with $\delta^2(\tau) \propto \tau$ sets in at longer lags $\tau \gg \tau_r$. It is worth emphasizing that these results hold for (overdamped) particles endowed with a self-propulsion of constant magnitude, so that, in the droplet system considered in this work, one can reasonably expect that the presence or features of its dynamical regimes will be influenced by the non-stationarity of the agents' activity \cite{peruani2007self,babel2014swimming,gutierrez2025time}---as well as, naturally, their mutual interactions. The explicit inclusion in the TSAMSD $\delta_{\mathcal{S}}^2(\tau,t_w)$ of the window running time $t_w$, alongside the lag time $\tau$ and species identity $\mathcal{S}$, aims at shedding light on these effects, tracking how the nature of the motion of the two droplet species evolves throughout the experiment as a consequence of the gradual decay of their self-propulsion.

Representative curves displaying the $\tau$ dependence of the $\delta_{\mathcal{S}}^2(\tau,t_w)$ for blue and red droplets at the beginning of each stage of the system's evolution---specifically, for times $t_w$ located in the mid of the colored vertical bars of Fig.~\ref{fig:order}---are reported in Fig.~\ref{fig:TAEMSD_stages}a-e. We note that the profiles corresponding to (the onset of) stages I-IV exhibit an evident crossover behavior in $\tau$, where for both species a single transition between a short and a long time-lag dynamic regime can be identified in Fig.~\ref{fig:TAEMSD_stages}a-d. On the other hand, the largely steady character of the system in stage V, see Fig.~\ref{fig:TAEMSD_stages}e, is such that some care is needed in the interpretation of the associated results. For this reason, the TSAMSD of stages I to early IV will, in the following, be analyzed separately from those ranging from late stage IV to V, IV marking the beginning of the system-wide ultimate decline in the self-propulsion of the agents.

To enable a quantitative analysis of the two distinct dynamical regimes that are observed in the system for non-vanishing activities, two characteristic time-frames were introduced, namely a short-lag regime, $\tau \in [0.1, 1]$~s, and a long-lag regime, $\tau \in [50, 300]$~s. Based on this decomposition, the family of TSAMSD curves was then fitted separately in each of these two time intervals \emph{via} a power-law relation of the form 
\begin{equation}
    \delta_{\mathcal{S}}^2(t_w,\tau) \sim K_{\mathcal{S}}^{dr}(t_w) \tau^{\alpha_{\mathcal{S}}^{dr}(t_w)}.
    \label{eq:power_law_fit}
\end{equation}
In Eq.~\ref{eq:power_law_fit}, for each dynamical regime $dr$ (short/long), each species $\mathcal{S}$ and at every window time $t_w$, the free parameters of the fit are $\alpha_{\mathcal{S}}^{dr}(t_w)$ and $K_{\mathcal{S}}^{dr}(t_w)$, respectively representing the short/long time-lags ``instantaneous'' dynamic scaling exponent and generalized diffusion coefficients of the droplets, where we underline that the quotes account for the fact that TSAMSD results were actually obtained by performing averages over a time window. 

\begin{figure*}[t]
    \centering
    \includegraphics[width=\textwidth]{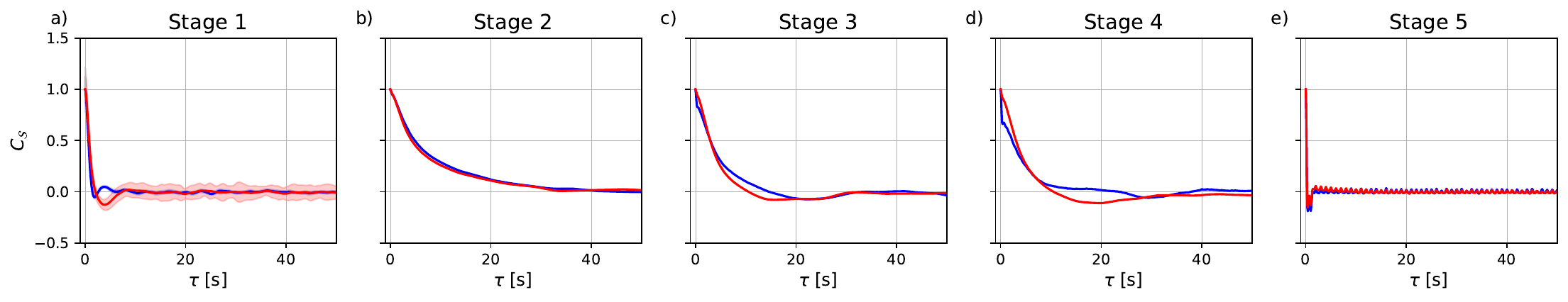}
    \caption{Velocity autocorrelation functions of the two droplet species, see Eq.~\ref{eq:VACF}. (a-e) Progression from left to right illustrates the $\tau$-dependence of the family of $C_{S}(t_w,\tau)$ VACF curves for the blue (blue line) and red (red line) species at the onset of each evolutionary stage of the system---that is, for window times $t_w$ located in the middle of each of the five colored vertical bars in Fig.~\ref{fig:order}}.
    \label{fig:vacf_evolution}
\end{figure*}

The results of the power-law fits in the short time-lag regime indicate that the dynamics of the droplets, from stage I to the onset of stage IV (see Fig.~\ref{fig:TAEMSD_stages}a-d), follows a quite robust $ \delta^2_{\mathcal{S}}(\tau,t_w) \propto \tau^{\alpha^{short}}$ scaling with $\alpha^{short} \simeq 2$ for both species and irrespective of the window running time $t_w$, see Fig.~\ref{fig:TAEMSD_stages}f. This is indicative of a short-time ballistic motion of the agents, which persists from the initial, high-activity phase of uncorrelated dynamics of the system up to the one in which the cluster of blue droplets is fully formed and surrounded by the still relatively disordered and motile red component. In the short-lag region, the main discrepancy between the TSAMSD curves in Fig.~\ref{fig:TAEMSD_stages}a-d across droplet color and evolutionary stage lies in the differences that occur between the intercepts of the $\delta^2_{\mathcal{B}}$ and $\delta^2_{\mathcal{R}}$ curves with the vertical axis in a log-log scale; these intercepts are linked to the short time-lag generalized diffusion coefficients $K_{\mathcal{S}}^{short}$ in Eq.~\ref{eq:power_law_fit}---a now formally counterintuitive naming given the presence of ballistic motion---whose variations over the course of the experiment, displayed in Fig.~\ref{fig:TAEMSD_stages}g, quite naturally follow those of the (square of the) agents' average speeds $V_{\mathcal{S}}(t_w)$ reported in Fig.~\ref{fig:order}c.

Specifically, as previously discussed, red droplets are faster than blue ones at the beginning of stage I, reflecting in $K_{\mathcal{R}}^{short}$ being larger than $K_{\mathcal{B}}^{short}$ and in the ballistic portion of the red TSAMSD curve lying above its blue counterpart in Fig.~\ref{fig:TAEMSD_stages}a. A gradual equalization of the speeds of the two species is then observed as the series of transient, mixed structures progressively emerge; when such parity is reached, we have $K_{\mathcal{R}}^{short}\approx K_{\mathcal{B}}^{short}$ and the blue and red TSAMSD curves are found to fully overlap over the whole range of analyzed lag times, as shown in Fig.~\ref{fig:TAEMSD_stages}b for the beginning of stage II. Consistent with the remainder of the metrics analyzed in this work, this suggests that the two droplet species contribute almost evenly in the generation of the metastable structures that appear for intermediate self-propulsion magnitudes. Subsequently, from stage II to early stage IV and with the emergence of the blue cluster, a steady decrease in the speed of all agents takes place, with the decline occurring more rapidly in the blue component. This results in a progressive and asymmetric decrease in $K_{\mathcal{R}}^{short}$, and $ K_{\mathcal{B}}^{short}$, and consequently in the intercepts of the TSAMSD curves of both species, leading the ballistic section of $\delta^2_{\mathcal{B}}$ to locate again under $\delta^2_{\mathcal{R}}$, see Fig.~\ref{fig:TAEMSD_stages}c~and~d.

As previously highlighted, increasing the time lag $\tau$ in the TSAMSD curves associated with the beginning of stages I-IV shown in Fig.~\ref{fig:TAEMSD_stages}a–d reveals a crossover behavior, with the agents entering the second dynamical regime that characterizes their motion over ``long'' timescales. The results of the power law fit via Eq.\ref{eq:power_law_fit} in this long-lag regime, presented in Fig.~\ref{fig:TAEMSD_stages}f,g, reveal that from early stage I to the onset of stage IV the two droplet species on average exhibit similar and somewhat mildly sub-diffusive dynamics at long time-lags, as witnessed by the corresponding scaling exponents $\alpha_{\mathcal{S}}$ fluctuating around a mean $\bar{\alpha}^{long}\lesssim 1$, see Fig.~\ref{fig:TAEMSD_stages}h. On the other hand, the long time-lag generalized diffusion coefficients $K_{\mathcal{S}}^{long}$ reported in Fig.~\ref{fig:TAEMSD_stages}i follow a temporal trend that is again qualitatively compatible with the one displayed by the droplet's mean speeds $V_{\mathcal{S}}$ in Fig.~\ref{fig:order}. More specifically, $K_{\mathcal{R}}^{long}$ is significantly higher than its blue-species counterpart at the beginning of the experiment and during the phase of uncoordinated motion, but rapidly decays throughout stage I, eventually becoming comparable to $K_{\mathcal{B}}^{long}$ as the series of transient structures appears---we remind that the red and blue TSAMSD profiles fully overlap over these timescales, see Fig.~\ref{fig:TAEMSD_stages}b. Subsequently, with the demixing transition between the two species that unfolds during stage II and reaches stability in stage III, both $K_{\mathcal{B}}^{long}$ and $K_{\mathcal{R}}^{long}$ continue to decrease, although the latter does so more slowly due to the higher residual activity of the red species, which is furthermore not trapped within the blue cluster.

We now proceed to examine the TSAMSD results for the two last stages of the system's evolution (late IV and V), during which, with the ultimate disappearance of the droplets' self-propulsion, the system enters its final, dynamically steady, and structurally ordered state. In this respect, the $\delta^2_{\mathcal{B}}$ and $\delta^2_{\mathcal{R}}$ profiles associated with (the beginning of) stage V reported in Fig.~\ref{fig:TAEMSD_stages}e reveal a more intricate dynamical picture; in particular, notable changes in the scaling properties of the TSAMSD at both short and long time lags are observed with respect to the case of higher activities, which appear to be further accompanied by the emergence of additional dynamical regimes at intermediate values of $10 \lesssim \tau \lesssim 50$~s.

Regarding the dynamics of the droplets on short time lags, an analysis of the scaling exponents associated with the $\delta^2_{\mathcal{S}}(\tau, t_w)$ curves reveals that, as the window time $t_w$ increases from the onset of stage IV, what was previously a ballistic motion of the agents at short lags gradually transitions to an approximately (super)diffusive behavior upon approaching stage V---first for the blue species and subsequently for the red one, see Fig.~\ref{fig:TAEMSD_stages}f. Conversely, the dynamical regime of both species at long time lags shifts from (sub)diffusive to ballistic in passing from IV to V, as evidenced by the rise in the corresponding scaling exponents to $\alpha^{long}\approx 2$ appreciable in Fig.~\ref{fig:TAEMSD_stages}h. This remodulation in the short and long time-lag dynamic properties of the system as it approaches its ``death'' phase can be readily interpreted; in fact, a largely steady condition is reached---see the agents' mean speeds in Fig.~\ref{fig:order}c and the short/long time lags generalized diffusion coefficients in Fig.~\ref{fig:TAEMSD_stages}g-i---while the red species' residual activity ultimately disappear along stage IV, with the fully-formed cluster of blue droplets being surrounded and caged by a chain of red ones. Due to their limited mobility, the extracted trajectories of the droplets get, for short time lags, increasingly dominated by the noise in the detection, which artificially renders the reconstructed dynamics of the agents diffusive-like on small timescales. The cluster as a whole (including the red species enclosing it), on the other hand, is still free to perform coherent, rigid‐body motion of very low speed: such a collective movement reflects in the generation of a ballistic scaling of the TSAMSD of the two species at long time lags, contrarily to the (sub)diffusive behavior that was instead observed for higher self-propulsion magnitudes. 

Finally, in addition to the changes in these two limiting cases, intermediate dynamical regimes appear to emerge in the TSAMSD of both species within the mid-$\tau$ range of $10 \lesssim \tau \lesssim 50$~s. These include plateaus in the droplets’ average displacement, indicative of a certain degree of arrested dynamics (see Fig.~\ref{fig:TAEMSD_stages}e). A detailed inspection of the ensemble of the TSAMSD profiles along stages IV and V reveals that the features of these intermediate regimes (such as the number of independent plateaus and their length) depend strictly on the time window $t_w$ (data not shown), and their thorough characterization will be the subject of future work.

We conclude this analysis of the system’s translational dynamic properties and their dependence on the activity level with a brief and qualitative overview of the velocity autocorrelation functions (VACF) of the agents, $C_{S}(t_w,\tau)$, defined in Eq.~\ref{eq:VACF}. Representative curves of the $\tau$ dependence of the VACFs at the beginning of each evolutionary stage I-V of the system (again, for times $t_w$ located in the middle of the colored vertical bars in Eq.~\ref{fig:order}) are presented in Fig.~\ref{fig:vacf_evolution} separately for the two components. Overall, in stage I and in the regime of the highest activity of the droplets, we observe a fast decorrelation of the velocity of both species, primarily as a consequence of the frequent collisions among the agents. The decorrelation time increases substantially in stages II (where the red and blue curves again fully overlap), III, and IV, and hence as the structural organization of the agents gradually appears---first in the form of the series of mixed, transient structures and then with the emergence of the cluster of blue droplets. Finally, in stage V, when the system is largely immobile, we observe a swift decay of the VACF, consistent with the fact that, in this regime, the estimation of the droplets' velocity is dominated by the noise in the detection.

\subsection{Turning angle distributions}

 We now turn our attention to the analysis of the droplets' \emph{rotational} diffusive features. Specifically, we consider the window-averaged distributions $P_{\mathcal{S}}(\Delta \theta,t_w)$,  $\mathcal{S}=\mathcal{B},\mathcal{R}$, defined in Eq. \ref{eq:turn_angl}, where $\Delta \theta \in (-\pi,\pi]$ is the turning angle, i.e. the angle between the versor of the velocity of an agent computed at two consecutive times.

 \begin{figure*}[t]
    \centering
    \includegraphics[width=\textwidth]{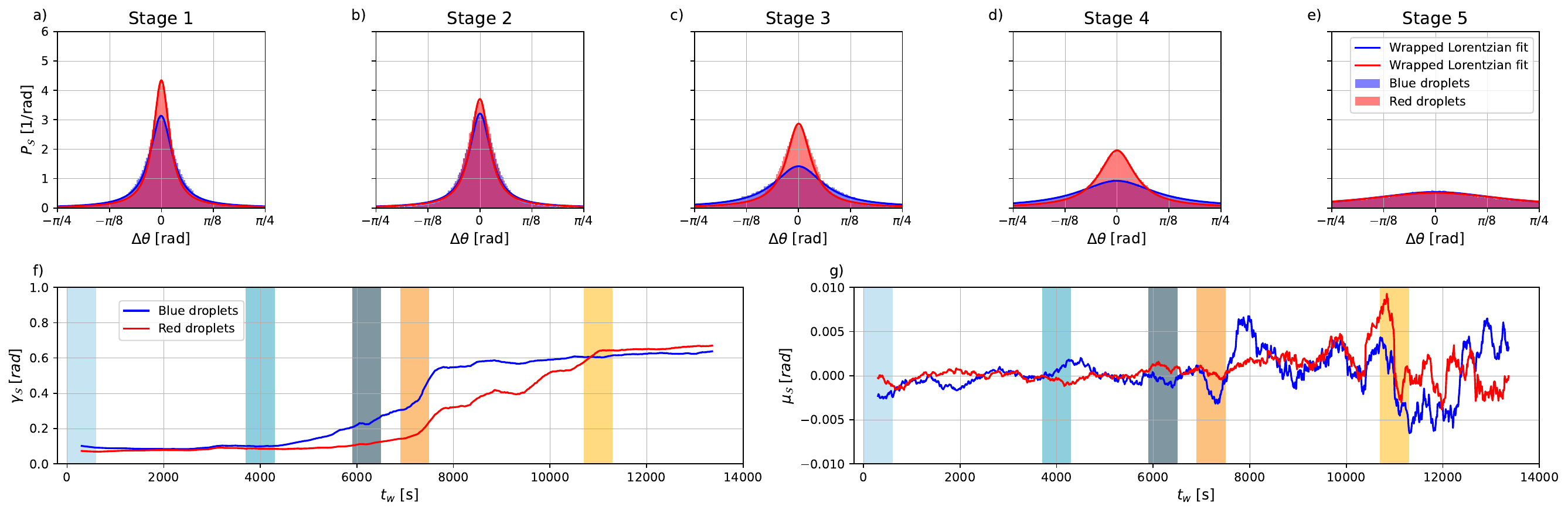}
    \caption{Turning angle distributions of the two droplet species, see Eq.~\ref{eq:turn_angl}. (a-e) Progression from left to right illustrates the $\Delta\theta$-dependence of the family of $P_{\mathcal{S}}(\Delta \theta,t_w)$ turning angle distributions for the blue (blue line) and red (red line) species at the onset of each evolutionary stage of the system---that is, for window times $t_w$ located in the middle of each of the five colored vertical bars in the plots below. (f-g) Temporal evolution of the scale factor $\gamma_{\mathcal{S}}(t_w)$ and of the mean $\mu_{\mathcal{S}}(t_w)$ of the blue and red droplet species resulting from the wrapped Lorentzian fit of the family of $P_{\mathcal{S}}(\Delta \theta,t_w)$ profiles, see Eq.~\ref{eq:wrapped_cauchy}. Colored vertical bars indicate the onset of each evolutionary stage.}
    \label{fig:turn_angles_stages}
\end{figure*}

A representative profile of $P_{\mathcal{S}}(\Delta \theta,t_w)$ for the blue and red droplets in each evolutionary stage of the system (I-V) is presented in Figs.~\ref{fig:turn_angles_stages}a-e, where, in all cases, $t_w$ was set to the initial one of the corresponding stage---that is, the center of the colored bars reported in Fig.~\ref{fig:order}. Regardless of the stage and the droplet species considered, a smooth dependence of the $P_{\mathcal{S}}$ on $\Delta \theta$ can be readily appreciated, with all the distributions being unimodal and centered around a vanishing turning angle. The extent of the droplets' \emph{deviations} from this preferred forward direction of motion, at the same time, critically depends on their species as well as on the specific evolutionary stage of the system, with the width of the profiles in Figs.~\ref{fig:turn_angles_stages}a-e displaying an overall increase as the self-propulsion decays.

To better quantify these features, a fit of the time-dependent ensemble of $P_{\mathcal{S}}(\Delta \theta,t_w)$ to functional forms that allow an intuitive interpretation of the data proves necessary, and the first, most natural choice is to employ normal distributions wrapped on the unit circle, approximating them \emph{via} von Mises functions \cite{mardia2009directional}; this attempt, however, returns generally inadequate results due to the relatively long tails exhibited by the $P_{\mathcal{S}}(\Delta \theta,t_w)$. In contrast, Figs.~\ref{fig:turn_angles_stages}a–e reveal that a good agreement, further consistent across the entire experiment, is achieved when each profile is fitted to a wrapped Cauchy distribution $f^W_\mathcal{S}(\Delta \theta,\mu_{\mathcal{S}}(t_w),\gamma_{\mathcal{S}}(t_w))$, with
\begin{equation}
 f^W_\mathcal{S}(\Delta \theta,\mu_{\mathcal{S}},\gamma_{\mathcal{S}})=\frac{1}{2\pi}\frac{\sinh(\gamma_{\mathcal{S}})}{\cosh(\gamma_{\mathcal{S}})-\cos(\Delta\theta-\mu_\mathcal{S})}.
 \label{eq:wrapped_cauchy}
\end{equation}

For each species and at every window time, the free parameters of the fit are $\mu_{\mathcal{S}}$ and $\gamma_{\mathcal{S}}$ in Eq.~\ref{eq:wrapped_cauchy}, respectively representing the mode and half-width at half-maximum of the Cauchy distribution $f_\mathcal{S}(\Delta\omega,\mu_{\mathcal{S}},\gamma_{\mathcal{S}})$ from which the wrapping on the unit circle is performed, where
\begin{equation}
f_\mathcal{S}(\Delta\omega,\mu_{\mathcal{S}},\gamma_{\mathcal{S}})=\frac{1}{\pi}\frac{\gamma_{\mathcal{S}}}{ \gamma_{\mathcal{S}}^2+ (\Delta\omega - \mu_{\mathcal{S}})^2},
\label{eq:cauchy}
\end{equation}
and $\Delta \theta = \Delta \omega \mod 2\pi$.

Relying on the good agreement between the data for $P_{\mathcal{S}}(\Delta \theta,t_w)$ and the corresponding fitted profiles, the evolution of the rotational diffusion properties of the droplets alongside the decay of their activity can now be investigated by analyzing the behavior, presented in Figs.~\ref{fig:turn_angles_stages}f and~g, of the set of time-dependent parameters $\mu_{\mathcal{S}}(t_w)$ and $\gamma_{\mathcal{S}}(t_w)$ in Eq.~\ref{eq:wrapped_cauchy}. For each species $\mathcal{S}$, the associated $\mu_{\mathcal{S}}(t_w)$ characterizes the ``instantaneous'' mean turning angle between two consecutive displacements of one of its droplets, where the quotes account for the fact that the $P_{\mathcal{S}}(\Delta \theta,t_w)$ profiles underlying the fits were obtained \emph{via} accumulating the distribution over a time window. Critically, Fig.~\ref{fig:turn_angles_stages}g reveals that both $\mu_{\mathcal{S}}(t_w)$ are compatible with zero throughout the whole experiment: this quantitatively confirms what was already hinted by the visual inspection of the reconstructed distributions, namely that droplets on average tend to maintain their direction of motion---that is, $\Delta\theta=0$. We note that the fluctuations of the $\mu_{\mathcal{S}}(t_w)$ around this value somewhat increase (albeit remaining extremely small) in stages IV and V, when the system becomes largely arrested and dynamic results are dominated, on short timescales, by the noise in the detection.

On the other hand, the scale factors $\gamma_{\mathcal{S}}(t_w)$ in Eq.~\ref{eq:wrapped_cauchy} control the breadth of the associated distributions, thus characterizing the extent to which each droplet species can deviate from purely forward motion; furthermore, they capture, through their explicit time dependence, if and how the magnitude of such deviations evolves with the progressive decay of the self-propulsion. In contrast to the case of the $\mu_{\mathcal{S}}(t_w)$, the scale factors $\gamma_{\mathcal{S}}(t_w)$ obtained from the fitting procedure and reported in Fig.~\ref{fig:turn_angles_stages}g highlight a non-trivial species- and time- dependence of the system’s rotational diffusive properties. Additionally, the observed trends again support the partition of the system’s evolution into the set of stages initially identified based solely on the droplets' velocity polarizations and hexatic order parameters.

More specifically, in the disordered state associated with the onset of stage I, both $\gamma_{\mathcal{B}}$ and $\gamma_{\mathcal{R}}$ are small ($\sim 0.1$~rad), reflecting how the strong self-propulsion characteristic of the early phase of the experiment leads to a quite markedly oriented motion for both species. Nonetheless, $\gamma_{\mathcal{B}}$ is slightly larger than its red component counterpart, indicating that the blue droplets, initially less active, can exhibit somewhat broader turning angles. The two scale factors gradually become comparable along stage I; this confirms, consistently with what was observed for the droplets' average speed, the associated polarizations, the TSAMSDs, and the VACFs, that the two species display similar dynamic (albeit not structural, \emph{vide infra}) properties in the series of transient structures that are generated on these timescales.

During stages II and III and as the activity of the droplets continues its decrease, a steady growth in $\gamma_\mathcal{B}$---and hence a smearing of the associated turning angle distribution---is observed, while the scale factor $\gamma_\mathcal{R}$ of the red component remains approximately constant to the value it had at the onset of the experiment and only displays a mild increase along III. This behavior is closely tied to the demixing transition that unfolds over such stages, culminating in the formation of the ordered cluster of blue droplets that gets increasingly caged by the red ones, see Figs.~\ref{fig:fig1} and \ref{fig:order}. The rise in $\gamma_\mathcal{B}$ reflects the fact that, alongside the growth of the blue cluster, the individual, self-propelled component of the motion of its constituent droplets fades, and is gradually replaced by the more randomly-oriented, collective dynamics of the cluster, driven by the asymmetric kicks it receives from the surrounding red species. This, by contrast, persists (although with decreasing strength) in its activity-driven, disordered motion, as other observables previously suggested and is now further witnessed by the consistency of $\gamma_\mathcal{R}$ throughout stages I–III.

With the ultimate disappearance of the residual activity of the red species along stage IV, the scale factor $\gamma_{\mathcal{R}}$ attains---albeit with a delay in the associated timescales---the value of $\gamma_{\mathcal{B}}$, indicating an increasingly negligible role of self-propulsion in the motion the agents. Eventually, saturation of both scale factors to $\sim 0.6$~rad is reached in the final, inactive stage (V) of the system, in which all droplets are essentially steady: here, as shown in Fig.~\ref{fig:turn_angles_stages}e, the turning angle distributions of the two species become largely indistinguishable and extremely broad, again signaling that the short-timescale dynamics reconstructed by our tracking pipeline gets dominated by the noise in the extraction of the droplets' center positions from the experimental videos.

\subsection{Distribution of droplets around a dimer}

To conclude our investigation, we shift our focus to the patterns that can be identified in the \emph{global} structural arrangement of the system through the set of (window-averaged) two-dimensional distributions $D_{\mathcal{S}\mathcal{S'}}(t_w)$ around a droplet \emph{dimer}. As outlined in Sec.~\ref{sec:metrics}, the methodology underlying these metrics draws inspiration from the work of Tanaka and collaborators \cite{doi:10.7566/JPSJ.86.101004}, who employed it in the context of a system which closely resembles the one examined in this study despite being mono-component. Specifically, in their analysis, Tanaka \emph{et al.} considered, for each configuration observed in an experiment, all the possible dimers formed by an agent $d_i$ and its nearest neighbor $n_i$, where $i = 1, \ldots, N$ and $N$ is the total number of droplets. They subsequently roto-translated the configuration so that $d_i$ was placed at the origin of the frame of reference and $n_i$ laid along the positive $x$-axis, repeating this operation for all possible choices of $d_i$; the $N$ replicas of each of the original snapshots resulting from this procedure were then employed to accumulate the average two-dimensional density of agents around a droplet dimer, where they further tracked the evolution of this metric along the course of the experiment \cite{doi:10.7566/JPSJ.86.101004}.

Applying this analysis workflow to our system requires us to explicitly account for its binary nature; accordingly, we evaluated four distinct dimer distributions, obtained by varying both the species composing the dimer and the species of the surrounding droplets whose average spatial arrangement was to be inspected. Specifically, we considered (\emph{i}) the distribution of blue droplets around a blue dimer, $D_{\mathcal{B}\mathcal{B}}$; (\emph{ii}) the distribution of red droplets around a red dimer, $D_{\mathcal{R}\mathcal{R}}$; (\emph{iii}) the distribution of red droplets around a blue/red dimer, $D_{\mathcal{B}\mathcal{R}}$, locating the origin of the frame of reference on the center of the blue droplet; and (\emph{iv}) the distribution of blue droplets around a red/blue dimer, $D_{\mathcal{R}\mathcal{B}}$, locating the origin of the frame of reference on the center of the red droplet. All these metrics were calculated within a window-based framework, $t_w$ being the running time lying at the center of the window over which distributions were accumulated. By examining the set of $D_{\mathcal{S}\mathcal{S'}}(t_w)$ distributions, our aim is to quantify the spatial correlations and potential preferential interactions occurring between droplets that belong to the same or different species on a global scale, thus complementing the more local information provided by the hexatic order parameters $\Psi_{6,\mathcal{S}}$ in Fig.~\ref{fig:order}b. Furthermore, and in analogy with Ref.~\cite{doi:10.7566/JPSJ.86.101004}, we will assess how these features are influenced by the amount of self-propulsion of the agents.

\begin{figure*}[ht]
    \centering
    \includegraphics[width=\textwidth]{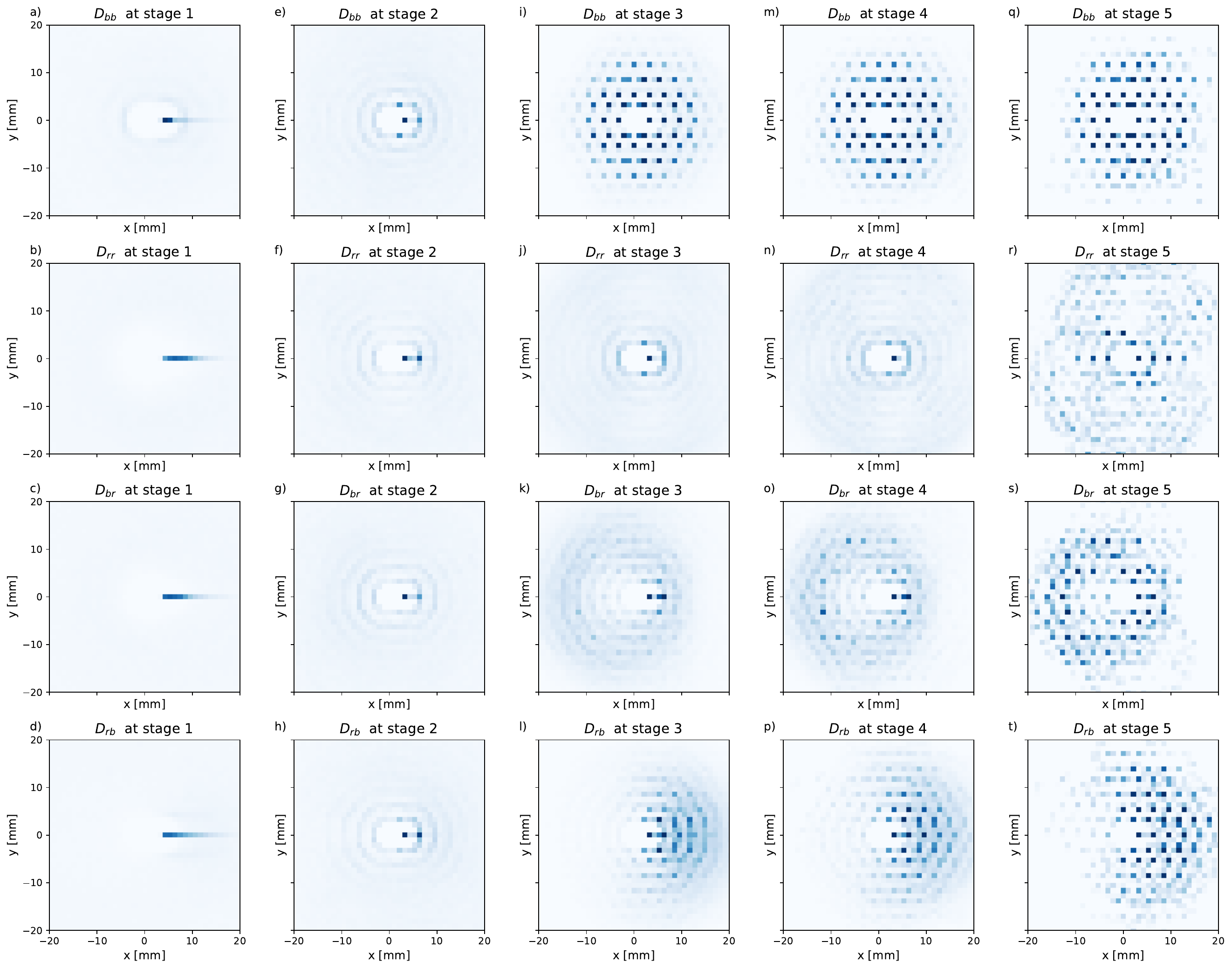}
    \caption{Heat maps of the dimer distributions of the two droplet species defined in Eq.~\ref{eq:dimer}. In each column, from top to bottom: distributions between blue droplets $D_{\mathcal{B}\mathcal{B}}$, red droplets $D_{\mathcal{R}\mathcal{R}}$, blue-red droplets $D_{\mathcal{B}\mathcal{R}}$, red-blue droplets $D_{\mathcal{R}\mathcal{B}}$. In each row, from left to right: illustrative heat maps of the dimer distributions at the onset of each evolutionary stage of the system. Darker color indicates higher density (arbitrary units).}
    \label{fig:dimer_distr}
\end{figure*}

Fig.~\ref{fig:dimer_distr} displays a representative profile for the blue/blue, red/red, blue/red, and red/blue dimer distributions across the five evolutionary stages (I–V) of the system. We underline that, in the plot, the set of $D_{\mathcal{S}\mathcal{S'}}(t_w)$ associated with each stage was extracted at the beginning of the latter---specifically, for running times $t_w$ located in the mid of each of the colored vertical bars reported in Fig~\ref{fig:order}. In analogy with the previously discussed metrics, it follows that the presented distributions are respectively characteristic of, see Fig~\ref{fig:fig1}: (\emph{i}) the system's initial phase of high self-propulsion and uncorrelated motion (early stage I); (\emph{ii}) the onset of the blue droplet cluster formation starting from the series of transient structures (early stage II); (\emph{iii}) the phase in which the fully-formed blue cluster is surrounded by the yet relatively disordered and active red species (the whole of stage III and the beginning of stage IV); and (iv) the system's final stationary phase of quasi-hexagonal order (stage V).

A bird's eye view of the distributions in Fig.~\ref{fig:dimer_distr} reveals the progressive increase in the regularity of the droplets' spatial arrangement that accompanies the decay of their self-propulsion: in fact, while largely homogeneous/diffused in the initial, high-activity phase, all the $D_{\mathcal{S}\mathcal{S'}}$ gradually evolve to develop sharp and localized peaks as the system progresses toward its final, steady regime of ``quasi-crystalline'' order. Key differences in the profiles can however be appreciated depending on the identity of the agents considered in the calculations, where we remind that, in a $D_{\mathcal{S}\mathcal{S'}}$ distribution, $\mathcal{S'}$ determines not only the species of the nearest neighbor of droplet $\mathcal{S}$ within the dimer, but also the species of the surrounding droplets whose average density is evaluated.

More specifically, at the beginning of the experiment and in the phase of highest self-propulsion of the system (early stage I), the distributions quantifying the red species' average spatial arrangement---that is, $D_{\mathcal{R}\mathcal{R}}$ and $D_{\mathcal{B}\mathcal{R}}$---are extremely homogeneous, see Fig~\ref{fig:dimer_distr}b,c. In addition to the excluded area associated with the impossibility of droplets' overlap, the only other visible feature in these profiles is a ``strip'' located along the positive x-axis, highlighting a quite delocalized probability to find the red neighbor of a red/red or blue/red dimer in a range of distances within $\sim [4.5 - 20]$~mm---equivalent to roughly $1$ to $4$ droplet diameters. Overall, this is indicative of the fact that, due to their initial high speed (see Fig.~\ref{fig:order}c), red droplets are completely unstructured in this regime, and on average distribute themselves uniformly throughout the Petri dish. The agents belonging to the slower blue species, on the other hand, exhibit a mild ``liquid-like'' structural order already at the onset of the experiment; this is evidenced, in the $D_{\mathcal{B}\mathcal{B}}$ profile reported in Fig.~\ref{fig:dimer_distr}a, by a diffuse hierarchy of concentric elliptical rings surrounding the dimer, which reflect the average coordination of additional neighboring blue droplets besides the nearest one---whose location probability along the x-axis is more neatly localized compared to the red species case. We note that this weak liquid-like ordering of the blue agents is not captured by the red counterpart due to its uniform spatial arrangement and high self-propulsion, as highlighted by the largely homogeneous red/blue dimer distribution $D_{\mathcal{R}\mathcal{B}}$ presented in Fig.~\ref{fig:dimer_distr}d.

Moving towards lower activities, Figs.~\ref{fig:dimer_distr}e-h report the dimer distribution characteristic of the initial phases of the demixing transition that, from the series of transient structures, leads to the generation of the cluster of blue droplets (beginning of stage II). Compared to the results obtained in the early stage I, we observe that the liquid-like ordering in shells of the blue agents around a blue/blue dimer, highlighted by the elliptical rings present in the $D_{\mathcal{B}\mathcal{B}}$ distribution, has become more neatly defined; furthermore, this kind of large-scale structural organization now extends also to the red species, with all the profiles in Fig.~\ref{fig:dimer_distr}e-h being largely compatible in terms of liquid-like coordination of neighbors. Consistent with the remainder of the analyzed metrics, this suggests that red and blue droplets play an analogous role in the series of transient structures generated on these timescales, likely because of their similar activities---see the agents' speed in Fig~\ref{fig:order}c. Subtle differences between the two species can instead be appreciated in terms of nearest neighbors' arrangement, with the somewhat delocalized strips that characterized their average location in stage I being now replaced by sharper peaks. As for the red species ($D_{\mathcal{R}\mathcal{R}}$ and $D_{\mathcal{B}\mathcal{R}}$ in Fig.~\ref{fig:dimer_distr}f,g), such peaks suggest a high probability of finding the red nearest neighbor at a distance of $1$ diameter from the reference (blue or red) droplet composing a dimer, and the next-to-nearest neighbor at $2$ diameters in the direction of the first neighbor, thus denoting that red droplets can arrange in short chain-like structures. In the case of the blue agents, these peaks are complemented by additional ones displaying hexagonal symmetry ($D_{\mathcal{B}\mathcal{B}}$ and $D_{\mathcal{R}\mathcal{B}}$ in Fig.~\ref{fig:dimer_distr}e,h), marking the appearance of what, later along this stage, will become the blue cluster.

The set of dimer distributions displayed in Fig.~\ref{fig:dimer_distr}i-p is instead representative of the phase of the system's evolution in which the cluster of blue droplets is fully formed, and is surrounded by the still relatively disordered and active red species driving its collective motion (stage III, Fig.~\ref{fig:dimer_distr}i-l, and early stage IV, Fig.~\ref{fig:dimer_distr}m-p). The structural regularity in the blue agents' spatial arrangement can be appreciated in the $D_{\mathcal{B}\mathcal{B}}$ profiles reported in Fig.~\ref{fig:dimer_distr}i and m; compared to their stage II counterpart, the fluid-like, diffused density rings of blue agents surrounding a blue/blue dimer have now disappeared, and the few localized peaks with hexagonal symmetry that previously only pertained to a droplet's nearest neighbors and marked the cluster seed now extend over the totality of the blue agents. As for the red species, the $D_{\mathcal{R}\mathcal{R}}$ distributions of red droplets around a red/red dimer shown in Fig.~\ref{fig:dimer_distr}j and n remain largely consistent with the stage II results, reflecting the still relatively active behavior and lack of regular and persistent spatial organization of the red component. Compared to Fig.~\ref{fig:dimer_distr}f, however, we note the appearance of a mild short-range hexatic ordering in the red nearest neighbors of a red droplet and, most importantly, the generation of a diffuse $8$-shaped structure in $D_{\mathcal{R}\mathcal{R}}$. Critically, the latter stems from the average circular arrangement of the red species around the blue cluster, where the effect of this caging manifests also in the emergence of an asymmetry in the blue/red and red/blue dimer distributions, see Fig.~\ref{fig:dimer_distr}k,l,o,p. $D_{\mathcal{B}\mathcal{R}}$, in fact, highlights a higher probability for a blue droplet to find a red agent in a circle oriented \emph{opposite} to the one of its red nearest neighbor, the profile remaining relatively diffuse as a consequence of the residual self-propulsion of the red species. Complementarily, $D_{\mathcal{R}\mathcal{B}}$ indicates and higher likelihood for a red droplet of finding a blue droplet in a direction that is \emph{forward} to that of its blue nearest neighbor, the more peaked nature of the distribution compared to $D_{\mathcal{B}\mathcal{R}}$ arising from the regularly ordered arrangement of the blue agents in the cluster. 

Finally, with the disappearance of the residual self-propulsion of the red species, the system takes a step toward its ultimate, steady, and ordered phase. The corresponding $D_{\mathcal{S}\mathcal{S'}}$ distributions, presented in Fig.~\ref{fig:dimer_distr}q-t (early stage V), remain largely compatible with the results obtained in stage III and IV; the only difference lies in the conversion of the residual diffused features present in the red/red, blue/red and red/blue profiles pertaining to the red species in sharper peaks, as a consequence of the globally vanished activity of the system.

\section{Discussion and conclusions}
\label{sec:conclusions}

 Artificial life research focuses on engineering synthetic active matter systems capable of exhibiting emergent behaviors akin to those characteristic of living organisms \cite{aguilar2014past}. Building on this vision, researchers in the field have devised chemical systems in which seemingly simple individual constituents are nonetheless capable of collectively generating intricate phenomena such as coherent motion and other life-like processes \cite{hanczyc2014droplets, loffler2023ecosystem,caschera2013oil,jancik2022swarming, krishna2024dynamic,cejkova2014dynamics,holler2018transport, holler2020autoselective,suzuki2016phototaxis, zarghami2019dual}; understanding how these features arise in controlled experimental setups may contribute to elucidating the physical principles governing their living analogues, with potentially significant scientific, technological and societal gains. 
 
 In this work, we investigated one such realization of artificial life, consisting of $50$ droplets of paraffin oil mixed with Ethyl Salicylate (ES) and suspended in an aqueous solution of Sodium Dodecyl Sulfate (SDS) within a Petri dish, where each active agent is roughly a millimeter in diameter. This system was recently proposed by Tanaka \emph{et al.} \cite{doi:10.7566/JPSJ.86.101004}, and in the case in which all droplets share the same chemical composition, it was shown to generate complex patterns of motion that resemble those observed in biological matter, patterns whose properties were further found to change along the course of the experiment due to the progressive decay in the droplets' self-propulsion. Increasing the complexity of the original setup, we here considered a situation in which half of the droplets are dyed red, while the other half are dyed blue; this different staining dramatically impacts the agents' features, with the red species having a stronger and longer-lasting activity than the blue one. Notably, the interplay of self-propulsions of different magnitudes and decay rates, hydrodynamic effects, and confinement results in a collective, multi-body, and heterogeneous behavior of the system throughout its evolution, in which, over roughly six hours, the agents pass from an initial regime of disordered motion to a static, species-demixed phase endowed with persistent structural order---a phenomenology consistently observed across all three independent experimental runs performed.

The challenges in the study of this system begin with acquiring quantitative and accurate data from its inherently noisy video recordings, a necessary step to translate the intuitive notion of ``emergent'' and ``life-like'' behavior observed in the experiments into a form suitable for a physics-based description, analysis, and characterization. To this end, we developed a system-tailored---yet general and adaptable---data extraction pipeline that takes the raw video footage as input and returns the high-resolution trajectory of each droplet along with other relevant features. We note that, for the scope of this work, droplets were treated as point-like particles located at their centers; while such an approach neglects, e.g., hydrodynamic effects and rotational inertia, it facilitates the systematic investigation of the system’s dynamical and structural properties. The trajectories obtained from this tracking pipeline were then processed through a series of metrics drawn from statistical physics, allowing us to quantify how the system's behavior evolves along the pathway it follows from an active, life-like state to a quiescent one of ``artificial death''.

The first step in the trajectory analysis has been the investigation of the system's degree of global structural and dynamical organization along the course of the experiment; to this end, we relied on two sets of collective variables, namely the velocity polarizations $\Phi_{\mathcal{S}}$ and hexatic order parameters $\Psi_{6,\mathcal{S}}$, where $\mathcal{S}$ marks the droplet species considered or the totality of the agents. The $\Phi_{\mathcal{S}}$ quantify the mutual alignment of droplet velocities and serve as a measure of the coherence in their motion \cite{PhysRevLett.75.1226}; being blind to the motion \emph{intensity}, these observables were interpreted in the light of the agents’ mean speeds $V_{\mathcal{S}}$. The hexatic order parameters, in contrast, assess how closely the neighbors of a droplet, depending on its species, are arranged at the vertices of a hexagon, and provide information on the emergence of structural ordering in the droplets' spatial arrangement \cite{jaster2004hexatic}.

The concurrent inspection of these observables allowed us to identify five distinct stages of the system's temporal evolution, further highlighting the different roles played by the red and blue droplet species due to their uneven activity levels and decay rates.
These stages are, respectively: (\emph{i}) an initial dynamically disordered and spatially homogeneous regime, followed by the emergence of a series of transient, loopy structures with mixed species composition; (\emph{ii}) a growth in the dynamical and structural ordering of the blue droplets, which display a faster self-propulsion decay and gradually organize in a cluster with hexagonal symmetry, expelling the red species; (\emph{iii}) a stable demixed phase of the two types of agents, where the still relatively disordered and active red component governs the blue cluster's collective dynamics; and eventually (\emph{iv}) a transition towards (\emph{v}) the final ``death'' phase, in which droplets are arranged in a essentially fixed, quasi-regular and phase-separated structure, with the cluster of blue agents being surrounded and caged by a chain of red ones. Notably, these stages are semi-quantitatively the same for all three experimental runs performed.

The system was subsequently investigated in terms of several additional observables, namely mean squared displacements, velocity autocorrelation functions, turning angle distributions, and density maps of droplets around a dimer. The analysis of these metrics and of their temporal dependence revealed a multifaceted picture of the system's behavior, offering insights that extend beyond those inferred solely from the order parameters.

The droplets' translational diffusive properties were found to remodulate along the course of the experiment and as the system's self-propulsion gradually decays. More specifically, in the early stages of the evolution and for high activity levels, both species exhibited ballistic motion of the agents at short time lags and a (mildly sub-)diffusive regime at longer times, accompanied by rapidly decaying velocity autocorrelations indicative of ``collisional'' dynamics. As time progresses and the activity of the droplets reaches intermediate levels, the ballistic and (sub-)diffusive character of their short- and long-time dynamics persists; still, long-time velocity autocorrelations emerge, reflecting sustained directional motion arising from the appearance of collective organization. Finally, once mobility has vanished and an ordered, largely steady cluster has fully formed (now comprising, albeit demixed, the entirety of the agents), the short-time lag dynamics of the individual droplets becomes increasingly dominated by the noise in the detection, while the long-lag behavior turns to ballistic due to the extremely slow, collective drift of the cluster.

Subsequently, we investigated the properties and time-dependence of the distributions $P_{\mathcal{S}}$ of turning angle between two consecutive displacements of a droplet. For each species, the associated $P_{\mathcal{S}}$ quantifies the directional coherence of the motion at the level of one of its agents. As shown in Fig.~\ref{fig:turn_angles_stages}, both such curves are compatible with a (wrapped) Cauchy law, with means consistently around zero throughout the whole experiment and widths that instead evolve over time. The distributions are initially rather narrow and indicative of a neatly forward-directed motion of the droplets; as activity gradually decays, both $P_{\mathcal{S}}$ distributions broaden---albeit differently for the two species---indicating that the agents’ displacements increasingly span a wider range of directions from one frame to the next.

The finding that the droplets' turning angle profiles are compatible with a Cauchy distribution represents an interesting result of this work, where we note that this angle differs from the one typically considered in theoretical descriptions of active matter---the focus in such cases being usually on the angle between two consecutive orientations of the self-propulsion vector \emph{alone} \cite{breoni2020active}. Whether the observed Cauchy-like distribution of the turning angle between what, in our case, are instead \emph{consecutive displacements} is a feature of the individual droplets or it emerges as a collective phenomenon, originating from the multi-body interactions among the agents, is the subject of ongoing investigation.

Lastly, we analyzed the droplets' structural arrangement relying on the method proposed by Tanaka \emph{et al.}~\cite{doi:10.7566/JPSJ.86.101004}, aligning trajectory frames relative to a reference dimer and calculating the average density of droplets around it, varying both the dimer identity and the species examined. The resulting maps revealed a steady progression of the system as self-propulsion gradually decays: from an initially roughly uniform distribution, to the formation of well-defined ``solvation shells'', and finally to hexatic order. Blue droplets attained structural regularity rapidly, while red droplets organized more slowly. Crucially, the asymmetry in the maps observed from intermediate activity levels through the ultimate system's ``death'' phase revealed the asymmetry in the droplets' behavior, with red agents preferentially arranging in a layer encircling the blue droplets.

In conclusion, the results of this work reach in two directions. First, we have introduced, in a known experimental active droplets system \cite{doi:10.7566/JPSJ.86.101004}, a small yet consequential element of novelty, namely the two dyes and the different activities they determine in the agents they stain; this led to the onset of emergent behavior hitherto unseen in this specific realization of artificial life, such as demixing and species-dependent spatial segregation. Three instances of the system under study have been investigated for an extended time frame, allowing us to verify not only the unfolding of qualitatively distinct stages of their evolution, but also the consistency of such stages across the different instances. Second, we have developed a general, efficient, and accurate tracking pipeline---which we make publicly available---that takes in the raw and extensive video recordings of the experiments and returns the detailed, high-resolution trajectory of each droplet, allowing us to carry out a quantitative investigation of the system's properties.

The results obtained pave the way for both fundamental investigations and practical developments. On the system side, the specific artificial life realization examined in this work can be broadened by playing with several parameters, including droplet number density, number and type of dyes (and thus agents' activity levels), and size and shape of the container, to investigate how these factors influence the system's emergent properties and its overall ``life-to-death'' cycle. Additionally, the unspecific functioning of the tracking and analysis pipeline lends itself to the study of greatly different systems, making it an enabling tool to boost the investigation of meso- to macro-scale active matter.

Last but not least, we observe that one of the most challenging aspects in the development of artificial life systems is the construction of accurate and predictive \textit{in silico} models, whose critical importance lies not only in the fundamental understanding they offer but also in the efficient exploration of parameter space they enable, thereby guiding the design of experimental artificial life realizations akin to the one investigated here. The modelling step has to rely on an accurate and informative body of \textit{quantitative} empirical data to identify the salient physical features of the system and faithfully embed them into a mathematical description: the protocol developed for and applied in this work, which allows the acquisition of high-resolution data and their intelligible interpretation, thus ushers in new opportunities for understanding artificial life, supporting its computer-aided investigation, and ultimately fostering the development of novel systems with tailored properties and, in the long term, genuinely programmable life-like behaviors.

\section{Acknowledgments}

This project has in part received funding from the European Union's Horizon Europe EIC 2023 Pathfinder Open programme under grant agreement No 101129734. Views and opinions expressed are however those of the author(s) only and do not necessarily reflect those of the European Union or European Innovation Council and SMEs Executive Agency (EISMEA). Neither the European Union nor the granting authority can be held responsible for them. R.J.G.L. was supported by Grant 62828 from the John Templeton Foundation. The opinions expressed in this publication are those of the author(s) and do not necessarily reflect the views of the John Templeton Foundation. 

\section{Data and software availability}
The datasets generated and analyzed during this study are available from the corresponding author upon reasonable request. All software used in this work, including the tracking and analysis pipeline, is available at \url{https://github.com/skandiz/DropleX}, under the MIT license. Detailed instructions for reproducing the analyses and figures in this paper are included in the repository.

\section{Author contributions}

RM, RP, and MH (equal contribution) conceived the study and proposed the work plan; RL, SH, and MH designed and performed the experiments; MS and RM designed the droplet tracking and trajectory analysis pipelines; MS implemented the pipelines; MS and RM carried out the data analyses; RM and MS, together with RP and MH, interpreted the results; RM and MS drafted the paper, supported by RP and RL. All authors reviewed the results and approved the final version of the manuscript.

\bibliographystyle{ieeetr}
\bibliography{main.bib}
\end{document}